\documentclass[11pt]{article}
\pdfoutput=1

\usepackage{jcappub}
\usepackage{float,comment,bm,xstring,placeins}

\def\be{\begin{equation}}
\def\ee{\end{equation}}
\def\ba#1\ea{\begin{align}#1\end{align}}
\newcommand{\vs}{\nonumber\\}

\renewcommand{\v}[1]{\bm{#1}}
\newcommand{\vx}{\v{x}}
\newcommand{\vy}{\v{y}}

\newcommand{\vk}{\v{k}}
\newcommand{\vp}{\v{p}}
\newcommand{\vq}{\v{q}}

\newcommand{\refeq}[1]{Eq.~(\ref{eq:#1})}          
\newcommand{\refeqs}[2]{Eqs.~(\ref{eq:#1})--(\ref{eq:#2})}          
\newcommand{\reffig}[1]{Figure~\ref{fig:#1}} 
\newcommand{\reffigs}[2]{figures~\ref{fig:#1}--\ref{fig:#2}}
\newcommand{\refFig}[1]{Figure~\ref{fig:#1}}
          
\newcommand{\refsec}[1]{section~\ref{sec:#1}}          
\newcommand{\refSec}[1]{Section~\ref{sec:#1}}          
\newcommand{\refapp}[1]{Appendix~\ref{app:#1}}

\newcommand{\Om}{\Omega_m}
\newcommand{\Ob}{\Omega_b}

\renewcommand{\d}{\delta}

\newcommand{\dm}{\delta_m} 
\newcommand{\Km}{\mathcal{K}} 

\def\Del{\mathcal{D}}
\def\Mpch{\,h^{-1}{\rm Mpc}}

\def\lapl{\nabla^2}
\newcommand{\blapl}{b_{\lapl\d}}
\newcommand{\btd}{b_{\rm td}}

\newcommand{\otd}{O_{\rm td}}
\def\mp{m_p}

\newcommand{\<}{\left\langle}
\renewcommand{\>}{\right\rangle}

\def\knl{k_\text{NL}}
\def\Plin{P_\text{L}}

\title{Beyond LIMD bias: a measurement of the complete set of third-order halo bias parameters}

\author[a]{Titouan~Lazeyras,}
\author[a]{Fabian~Schmidt}
\affiliation[a]{Max-Planck-Institut f\"ur Astrophysik, Karl-Schwarzschild-Str. 1, 85748 Garching, Germany}
\emailAdd{titouan@mpa-garching.mpg.de} 
\emailAdd{fabians@mpa-garching.mpg.de}
\abstract{We present direct measurements of cubic bias parameters of dark matter halos from the halo-matter-matter-matter trispectrum.  We measure this statistic efficiently by cross-correlating the halo field measured in N-body simulations with specific third-order nonlocal transformations of the initial density field in the same simulation. Together with the recent Ref.~\cite{Abidi:2018eyd}, these are the first measurements of halo bias using the four-point function that have been reported to date. We also obtain constraints on the quadratic bias parameters. For all individual cubic parameters involving the tidal field $\Km_{ij}$, we find broad consistency with the prediction of the Lagrangian local-in-matter-density ansatz, with some indications of a positive Lagrangian coefficient $\btd^L$ multiplying the time derivative of $\Km_{ij}$. For the quadratic tidal bias ($b_{K^2}$), we obtain a significant detection of a negative Lagrangian tidal bias. }
\begin{document}

\maketitle
\flushbottom

\section{Introduction}
\label{sec:intro}

The large-scale distribution of dark matter halos is one of the key
ingredients of the theoretical description of large-scale structure (LSS).  
Since it is now well established that most observed tracers of LSS, such as galaxies, reside in halos (see e.g. \cite{Brainerd:1995da}),
the statistics of halos determine those of galaxies on large scales.  
In the context of perturbation theory, the statistics of halos are written in terms of bias parameters multiplying operators $O$ constructed out of the matter density field $\dm$ and the tidal field $\Km_{ij}$ (see \cite{biasreview} for a recent review)
\be
\d_h(\vx,\tau) = \sum_O b^E_O(\tau) O(\vx,\tau) \, ,
\label{eq:biasexp}
\ee
where $\d_h$ is the fractional number density perturbation of a given halo sample and $b_O^E$ is the bias parameter corresponding to the operator $O$. The superscript $E$ stands for Eulerian, since we are describing the statistics of the evolved (late-time) halo density field in terms of the evolved density and tidal fields. Since we will mainly focus on these parameters, we will drop the superscript $E$ throughout the paper. In contrast, we will explicitly use the notation $b_O^L$ to refer to Lagrangian (early time) bias parameters, that is, the parameters appearing in the expansion relating halos traced back to the initial conditions to the linearly extrapolated initial density field and tidal field.

The contributions in the general perturbative bias expansion \refeq{biasexp} can be classified in terms of the number of spatial derivatives acting on each instance of the gravitational potential appearing in the operators. The leading terms on large scales are those which involve exactly two spatial derivatives on each instance of the potential (here, we count $\partial_i\partial_j/\nabla^2$ as zero net derivatives). These constitute the leading local gravitational observables, and, following \cite{biasreview}, we consequently call this class the \emph{local bias} expansion. In particular, this class contains powers of the density field $\dm^n$ and tidal field $(\Km_{ij})^l$, as well as combinations of the two, and convective time derivatives of the tidal field \cite{senatore:2015,MSZ}. Note that, in the previous literature, terms involving the tidal field have often been referred to as ``nonlocal bias.'' However, since the tidal field is clearly a local observable \cite{baldauf/etal:2011,CFCpaper1}, it appears appropriate to include it in the class of local bias.
In contrast, following \cite{biasreview}, we will refer to the subclass of terms involving powers of the density field $\dm^n$, often referred to as ``local bias'', as \emph{local-in-matter-density (LIMD)} bias. One often-adopted ansatz is to assume that the halo bias expansion in Lagrangian space, i.e. in the initial conditions, only involves powers of the density field. We refer to this as the \emph{Lagrangian LIMD (LLIMD)} ansatz (often referred to as ``local Lagrangian'' or ``coevolution'' ansatz in the literature), and will compare our results with this assumption.

The second important class of contributions to \refeq{biasexp} involves more than two spatial derivatives on the gravitational potential. One example is a term $\nabla^2 \dm$. The key differences to the local bias contributions is that, first, the higher-derivative contributions are suppressed on large scales; for example, $\nabla^2\dm(\vx)$ becomes $-k^2 \dm(\vk)$ in Fourier space. Second, their coefficients are dimensionful; for example, $[b_{\nabla^2\d}] = \text{Mpc}^2$. Thus, their amplitude involves an additional spatial scale $R$. For dark matter halos, one expects this scale to be of order the halo Lagrangian radius (e.g., \cite{mcdonald/roy}). 

Since they are suppressed on large scales, higher-derivative bias contributes to the next-to-leading order correction to statistics, such as the 1-loop contribution to the halo power spectrum. Usually, the higher-derivative contributions are degenerate in shape with higher-order local bias contributions which also enter at next-to-leading order \cite{assassi/etal,angulo/etal}; for example, second- and third-order local bias terms appear in the 1-loop halo power spectrum. Hence, the higher-order local bias parameters are constrained most robustly by measuring higher $n$-point functions in the large-scale limit, rather than relying on the 1-loop halo power spectrum, for example.

Currently the LIMD parameters $b_n \equiv n!\, b_{\d^n}$ ($n\geq1$) are the most studied and have been measured up to $b_4$ in a variety of manners (see e.g. \cite{Matarrese:1997sk, Angulo:2007, guo/jing:2009, tinker/etal:2010,pollack/smith/porciani:2012, Saito:2014qha,Lazeyras:2015, Li:2015, Baldauf:2015,hoffmann/etal:2015,Hoffmann:2016omy} and references therein, as well as section~4.5 of \cite{biasreview} for a more exhaustive list). These methods include, but are not limited to, moments and scatter-plot methods \cite{Angulo:2007,manera/gaztanaga:2011,castorina/paranjape/etal:2016}, the separate-universe technique \cite{Lazeyras:2015,Li:2015,Baldauf:2015} (see e.g. \cite{McDonald:2001fe, Wagner:2014} for details about this technique), fits to the halo power spectrum and bispectrum \cite{guo/jing:2009,pollack/smith/porciani:2012} or the halo 3-point function \cite{hoffmann/etal:2015} , and correlators of operators constructed out of the squared density and tidal fields \cite{Schmittfull:2014tca} (the latter essentially measures the bispectrum as well, as we will see). The parameter $b_{K^2}$ has also been measured from the tree-level bispectrum in \cite{chan/scoccimarro/sheth:2012,baldauf/etal:2012,Saito:2014qha,angulo/etal}, the Lagrangian bispectrum \cite{sheth/chan/scoccimarro:2012}, the 3-point function \cite{bel/hoffmann/gaztanaga:15}, and from Lagrangian moments-based measurements \cite{castorina/paranjape/etal:2016, Modi:2016dah}. Some disagreement has been found between  \cite{Saito:2014qha} and \cite{Modi:2016dah} in the results for $b_{K^2}$. Finally, the parameter $b_{3\rm nl}$, a certain combination of quadratic and cubic tidal biases, has also been constrained from the 1-loop halo-matter power spectrum in \cite{Saito:2014qha}, but, following our discussion, it is degenerate with the higher-derivative bias, which was set to zero in that reference.

The goal of the present paper is to measure all the cubic bias terms, which are: $b_3 = 6 b_{\d^3}$, $\btd$, $b_{K^3}$, and $b_{\d K^2}$. The leading statistic to which these contribute is the four-point function (trispectrum). In order to measure them, we generalize a technique proposed by Ref.~\cite{Schmittfull:2014tca} to measure the relevant trispectrum contributions efficiently. This technique allows us to measure all the cubic bias parameters at once. Together with the recent Ref.~\cite{Abidi:2018eyd}, these are the first measurements of halo bias using the four-point function that have been reported to date. We further use the analogous technique for the bispectrum to measure $b_1$, $b_2$ and $b_{K^2}$, to cross-check our results obtained from the trispectrum, and to compare with previous measurements. 

The very similar study of \cite{Abidi:2018eyd} came out shortly after this paper appeared on the arXiv preprint server. While they use the same technique to obtain cubic order bias parameters from the trispectrum, some details such as higher-order corrections are treated differently. Overall, our results are in good agreement with theirs.

This paper is organised as follows: in \refsec{triest} we present our estimator for the trispectrum and show how to obtain the bias parameters from it. \refSec{sims} describes our set of simulations, shortly explains the halo finding procedure, and gives details on the actual procedure to measure the bias parameters. \refSec{predictions} reviews previous measurements and theoretical predictions for the parameters. Finally, \refsec{results} presents and discusses our results. We conclude in \refsec{concl}. The appendices contain details on the calculations (\refapp{biasexp} and \ref{app:subtraction}), higher-order corrections (\refapp{HO}), convergence tests and cross-checks of our results (\refapp{convergence} and \ref{app:check}), and covariance matrix of our results (\refapp{covariance}). We adopt the same cosmology as in \cite{Lazeyras:2015} (hereafter L15), i.e. a flat $\Lambda {\rm CDM}$ cosmology with $\Om = 0.27$, $h = 0.7$, $\Ob h^2=0.023$, $n_s=0.95$, and $\mathcal{A}_s = 2.2 \cdot 10^{-9}$. 

\section{Estimating cubic local bias from the trispectrum}
\label{sec:triest} 

We present here our estimator for the trispectrum and how it is used to measure the bias parameters. The estimator is based on the same idea as was first introduced in \cite{Schmittfull:2014tca}, which we briefly review in the first part of this section. We refer the reader to their paper for more details. Throughout the entire section we drop time arguments for clarity; the results can be applied at any redshift.

\subsection{Warmup: the squared-field method}
\label{sec:squaredfields}

Consider the halo density field at second order in perturbation theory:
\be
\d_h^{(2)}(\vx) = b_1 \d^{(2)}(\vx) + \frac12 b_2 \left[ \d^2(\vx) - \< \d^2 \>\right] + b_{K^2} \left[ K^2(\vx) - \< K^2 \> \right]\,,
\label{eq:dh2}
\ee
where the superscript $(n)$ indicates the order in perturbation theory and we drop the superscript $(1)$ for the linear fields for simplicity. The second-order density field is given by (see \refapp{biasexp})
\ba
\d^{(2)}&= \frac{17}{21} \d^2 + \frac27 (K_{ij})^2 - s^i\partial_i \d\,.
\label{eq:dm2}
\ea
The linear tidal field $K_{ij}$ and linear Lagrangian displacement are given respectively by
\ba
K_{ij} & = \mathcal{D}_{ij} \d = \left[\frac{\partial_i\partial_j}{\lapl}-\frac13\d^K_{ij}\right]\d \, ,
\label{eq:kij} \\
s^i & = - \frac{\partial_i}{\lapl}\d\,,
\label{eq:si}
\ea
where $\d^K_{ij}$ denotes the Kronecker symbol. 
We are interested in measuring the second-order bias parameters $b_2$ and $b_{K^2}$. The leading statistic in which the second-order halo field appears is the bispectrum, the simplest of which is the halo-matter-matter bispectrum,
\be
\< \d_h(\vk) \d_m(\vp_1)\d_m(\vp_2) \>'\,,
\label{eq:Bhmm}
\ee
where $\d_m(\vp)$ is the evolved fractional matter density perturbation. Here and throughout, a prime on a correlator denotes that the momentum-conserving Dirac delta is to be dropped, $(2\pi)^3 \d_D(\vk+\vp_1+\vp_2)$ in the present case. 
Since we have access to both the linear (initial) and nonlinear (evolved) density fields in the simulations, we can simply remove all contributions due to the nonlinearity of matter contained in $\d_m(\vp_i)$ in \refeq{Bhmm}, by instead considering the correlator
\be
\< \d_h(\vk) \d(\vp_1)\d(\vp_2) \>'\,.
\label{eq:BhmmL}
\ee
Following \cite{Schmittfull:2014tca}, we can compress the information in the three-dimensional phase space of the full bispectrum into a set of two-point correlations:
\be
\< \d_h(\vk) O^{(2)}[\d_R](\vk') \>'\,,
\label{eq:sqfield}
\ee
where $\d_R(\vk)$ denotes the linear density field smoothed on a scale $R$, and the quadratic operators are given by
\ba
O^{(2)}[\d_R](\vk) =\:& \int d^3\vx\: O^{(2)}[\d_R](\vx) e^{-i\vk\cdot\vx}\,,\vs
O^{(2)}[\d_R](\vx) \in\:& \Big\{ \d_R^2(\vx) - \< \d_R^2\>,\
(K_{ij,R})^2(\vx) - \< (K_{ij,R})^2 \>,\
s_R^i(\vx)\partial_i \d_R(\vx)
\Big\}\,.
\label{eq:O2dR}
\ea
That is, we cross-correlate the halo density field with the square of the linear density field and tidal field, and the displacement term appearing in $\d^{(2)}$, where, in all cases, the quadratic operators are constructed from the \emph{smoothed} linear density field $\d_R$. It is then clear that the correlators \refeq{sqfield} correspond to specific integrals of the halo-matter-matter bispectrum \refeq{BhmmL} over $\vp_1,\vp_2$. Inserting \refeqs{dh2}{dm2} into \refeq{sqfield},
we see that this cross-correlation becomes
\ba
\< \d_h(\vk)  O^{(2)}[\d_R](\vk') \>' =\:& \sum_{O'=\d^2, K^2,s^i\partial_i\d} c^{(2)}_{O'} M_{OO'}^{(2)}(k)\,,\vs
M_{OO'}^{(2)}(k) =\:& \< O^{(2)}[\d_R](\vk') O'^{(2)}(\vk) \>'\,,
\label{eq:sqfieldbias}
\ea
where the coefficient vector is given by (see \refapp{biasexp})
\be
\v{c}^{(2)} = \left(\begin{array}{c}
  b_2/2 + (17/21) b_1  \\
  b_{K^2} + (2/7) b_1\\
  - b_1 
  \end{array}
  \right)\,,
  \label{eq:Csq}
\ee
which contains the desired bias parameters $b_2$, $b_{K^2}$, as well as $b_1$.
In \refeq{sqfieldbias}, $O^{(2)}$ are constructed in the same way as $O^{(2)}[\d_R]$ [\refeq{O2dR}], but from the unsmoothed linear density field. 
This result is valid as long as $R$ and $1/k$ are sufficiently large, so that the correlator \refeq{sqfield} is accurately described by second-order perturbation theory. Then, $M_{OO'}^{(2)}(k)$ is given by a convolution integral over two linear power spectra, weighted by the Fourier-space kernels corresponding to the operators $\d^2,\,(K_{ij})^2,\,s^i\partial_i\d$ \cite{Schmittfull:2014tca}. However, we do not need these analytical expressions here, as $M_{OO'}^{(2)}$ can be directly evaluated on the simulations. 

The procedure to measure second-order halo bias now simply becomes:
\begin{itemize}
\item Construct the quadratic fields $O^{(2)}$ and $O^{(2)}[\d_R]$ using the linearly extrapolated initial density field used in the given simulation. This can be done efficiently on a grid by making use of fast Fourier transforms (FFT). Specifically, spatial derivatives and nonlocal operators such as $1/\nabla^2$ are applied in Fourier space, while products are taken in real space.
\item Measure the cross-power spectra between the halo field $\d_h(\vk)$ and the operators $O^{(2)}[\d_R](\vk)$, and the cross-power spectra of $O^{(2)}(\vk)$ and $O^{(2)}[\d_R](\vk)$. The latter yield $M_{OO'}^{(2)}(k)$.
\item Estimate the bias parameters by solving \refeq{sqfieldbias} for $\v{c}^{(2)}$.
\end{itemize}
In the following sections, we will provide more details on how these steps are implemented. 

Let us now briefly list the key differences between this work and \cite{Schmittfull:2014tca}. First, we construct our operators $O^{(2)}[\d_R]$ from the linear, rather than evolved matter density field used in \cite{Schmittfull:2014tca}. Second, rather than using the analytical expression for the ensemble average, we estimate the operator cross-power spectra $M_{OO'}^{(2)}(k)$ from the same realization of the initial density field. This is expected to further suppress cosmic variance in the estimated parameters $c_O^{(2)}$.

Finally, while Ref.~\cite{Schmittfull:2014tca} considered quadratic operators as an efficient means to measure the halo bispectrum, as we have just described, we will go to cubic order in order to measure the halo trispectrum. This is described next.

\subsection{Cubed-field method}

Consider the halo-(matter)$^3$ cross-trispectrum,
\be
\< \d_h(\vk) \d_m(\vp_1)\d_m(\vp_2)\d_m(\vp_3) \>_c\,,
\ee
where the subscript $c$ denotes the connected part of the four-point function. All cubic bias terms contribute to this statistic at tree level, in addition to the quadratic and cubic operators in the nonlinear matter density. Since we have access to the linear density field in the simulations, we can remove the second contribution by using, in analogy to \refeq{BhmmL},
\be
\< \d_h(\vk) \d(\vp_1)\d(\vp_2)\d(\vp_3) \>_c\,.
\ee
Since we are interested in cubic bias specifically, we can further simplify the statistic by subtracting the evolved matter density field, multiplied by the linear bias:
\be
\< \left[\d_h(\vk) - b_1 \d_m(\vk) \right] \d(\vp_1)\d(\vp_2)\d(\vp_3) \>_c\,.
\label{eq:ThmmmL}
\ee
$b_1$ can be measured for example from the large-scale halo-matter cross power spectrum (e.g., \cite{tinker/etal:2010}), or using the separate-universe technique \cite{Lazeyras:2015,Baldauf:2015,Li:2015}.
Note that, unlike the case for the bispectrum, there are disconnected lower-order contributions to the 4-point function, which we need to remove, as they do not involve cubic bias terms. Some of these are removed by subtracting the linear bias contribution multiplied by the \emph{nonlinear} density field. In addition, this subtraction removes contributions to the 6-point function from $\d^{(3)}$. 
Note that the lowest-order quadratic terms in $\delta_h$, while not of interest here, do not need to be subtracted, as they lead to 5-point functions which vanish. We include the quadratic bias contributions evaluated at third order, which contribute at leading order to the trispectrum, in our model.

Now, instead of attempting to measure the trispectrum \refeq{ThmmmL} for all possible
configurations in its six-dimensional phase space, one can again compress the information into a set of power 
spectrum-like quantities, by cross correlating $\d_h$ with \emph{cubic} operators $O^{(3)}[\d_R]$ constructed out of the smoothed linear density field on a scale $R$. As mentioned above, these can be constructed efficiently on a grid by going back and forth between real- and Fourier-space. We assume throughout that the mean of all operators has been subtracted, such that $\< O^{(3)}[\d_R](\vx) \> =0$ [in analogy to \refeq{O2dR}]. We will perform a further subtraction that removes the disconnected contributions below.

Paralleling the quadratic case discussed above, this measurement yields a
linear combination of operator cross-power spectra, multiplied by linear combinations, denoted as $c_O^{(3)}$, of the desired cubic and lower-order bias parameters:
\ba
\< \left[\d_h(\vk) - b_1 \dm(\vk) \right] O^{(3)}[\d_R](\vk') \>'  
= \sum_{O'} c^{(3)}_{O'} M_{OO'}^{(3)}(k)\,,
\label{eq:dhO3}
\ea
where
\ba
M_{OO'}^{(3)}(k) \equiv \< O^{(3)}[\d_R](\vk') O'^{(3)}[\d](\vk) \>' \,,
\label{eq:MOO}
\ea
and the vector of cubic operators is 
\be
\v{O}^{(3)} = \Big( \d^3,\, \d K^2,\, K^3,\, O_{\rm td},\, s^i\partial_i (\d^2),\,  s^i\partial_i (K^2)\Big)^\top\,,
\label{eq:operatorlist}
\ee
as is shown in \refapp{biasexp}. Here,
\ba
O_{\rm td} & = \frac{8}{21}K^{ij}\mathcal{D}_{ij}\left[\d^2-\frac32 K^2\right],
\label{eq:otd}
\ea
and $K^2=K_{ij}K^{ij}$, $K^3 = K_{ij} K^j_{\  l} K^{li}$. The displacement field is given by \refeq{si}. 
Each of the operators in \refeq{operatorlist} is cubic in linear fields.
The operators in \refeq{operatorlist}, when correlated with the halo density field, in general lead to lower-order, disconnected contributions to the 4-point function. Further, since we construct the operators as products in real space, there are zero-lag contributions to the correlators of the operators among themselves that appear on the right-hand side of \refeq{dhO3}. In the renormalized bias expansion, these contributions are removed by counter-terms. That is, we should employ the renormalized operators $[O^{[3]}]$ in \refeqs{dhO3}{MOO}. Since the cubic operators are constructed from the linear density field, this renormalization is in fact very simple. As shown in \refapp{subtraction}, both of these sets of unwanted contributions can be removed simultaneously by including the leading counter-terms to the bare operators, which are given by:
\ba
\v{O}^{(3)}(\vx) \to\:& [\v{O}^{(3)}(\vx)] = \v{O}^{(3)}(\vx) - \v{n}_O \< \d^2 \> \d(\vx)\, , \vs
\mbox{where}\quad \v{n}_O =\:& \Big( 3,\, 1,\, 0,\, 0,\, 2,\, 0 \Big)\,.
\label{eq:subtr}
\ea
Here, $\< \d^2 \>$ is the variance of the density field from which the operators are constructed, and $\d(\vx)$ is the same density field. These relations are derived in \refapp{subtraction}. This renormalization is analogous to the orthogonalization procedure described in \cite{Abidi:2018eyd}. However, the procedures differ in detail. In particular, we subtract a single term as written in \refeq{subtr}, while Ref.~\cite{Abidi:2018eyd} perform subtractions in Fourier space for each $k$ value individually.

Note that the coefficients $c^{(3)}_O$ contain contributions from $b_2$, $b_{K^2}$, since the halo density field at third order also contains the quadratic operators $\d_m^2, \Km^2$ evaluated at that order. In particular, the displacement terms, the last two operators in \refeq{operatorlist}, are multiplied by $-b_2/2$ and $-b_{K^2}$, respectively. This allows for important cross-checks. Specifically, as shown in \refapp{biasexp}, the set of coefficients $c_O$ is given by
\be
\v{c} := \v{c}^{(3)} = \{ c_O \}_{O^{(3)}} = \left(\begin{array}{c}
  b_3/6 + (17/21) b_2 \\
  b_{\d K^2} + (2/7) b_2 + (4/3) b_{K^2} \\
  b_{K^3} + 2 b_{K^2}  \\
  b_{\rm td} + (5/2) b_{K^2} \\
  - b_2/2 \\
  - b_{K^2}
  \end{array}
  \right)\,.
\label{eq:C}
\ee

Here and in the following, we drop the superscript $(3)$ as well as the brackets, as we are only dealing with renormalized cubic operators throughout the main text. Again, as in the quadratic case, $M_{OO'}$ denotes the cross-correlation of the \emph{unsmoothed renormalized operator} $O'$ with the renormalized operator $O$ \emph{constructed from the smoothed linear field $\d_R$}.
Thus, it is a specific scalar product between the operators $O$ and $O'$;
note that $M_{OO'}$ is not symmetric.
In the following, we will assume that the smoothing scale $R$ as well as wavenumbers $k$ are on sufficiently large scales so that the tree-level trispectrum is sufficient to describe the correlators in \refeq{dhO3}.

\subsection{Bias estimator}
\label{sec:biasest}

We now turn to the cubic bias estimator. We define the vector $\v{H}(k)$ of binned cross-power spectra of halos with these operators as defined in \refeq{dhO3}:
\be
H_O(k) = \sum_{k-\Delta k \leq |\vk| \leq k +\Delta k} \< \left[\d_h(\vk) - b_1 \d_m(\vk)\right] O[\d_R](\vk') \>'\,.
\ee
Using \refeq{dhO3}, this vector becomes
\be
\v{H}(k) = \v{M}(k) \cdot \v{c}\,,
\label{eq:biasestimator}
\ee
where $\v{c}$ contains the combinations of bias parameters given in \refeq{C}, and $\v{M} = \{ M_{OO'} \}$ is the matrix of operator cross spectra defined in \refeq{MOO}. 

We can then immediately construct the estimator for the vector of bias coefficients at any fixed $k$,
\be
\v{C}(k) = \v{M}^{-1}(k) \cdot \v{H}(k)\,.
\label{eq:biasestimatork}
\ee
Assuming that the smoothing scale $R$ is sufficiently large, $\v{C}(k)$ asymptotes to the scale-independent constant vector $\v{c}$ at sufficiently low $k$. The leading correction due to higher-order contributions can be approximated by a quadratic dependence on $k$:
\be
\v{C}(k)=\v{c}+\v{A} k^2 \, , \mbox{ and} \, \lim_{k\to 0} \v{C}(k) = \v{c}\,.
\ee
Thus, in this regime, one can combine the bias parameters from different $k$ bins. Further, if error estimates are available for $\v{C}(k)$ as a function of $k$, the estimates from different wavenumber bins can be weighted optimally, leading to an optimal estimator (at leading order) for the cubic bias parameters, as can be shown in analogy to the results of \cite{Schmittfull:2014tca}.

\section{Simulations and halo finding}
\label{sec:sims}

In this section, we shortly present the details of our set of simulations, and a quick outline of the halo finding procedure. We also detail the exact measurement procedure.

We use two sets of gravity-only simulations which were run with the cosmological N-body code GADGET-2 \cite{Springel:2005}. The first one has a box length $L=500 \Mpch$ with $N=512^3$ particles yielding a mass resolution $\mp=7\cdot 10^{10} h^{-1} M_\odot$. We ran 48 realisations of this set and refer to it as L500. In addition, we use two realisations of a larger box simulation with $L=2400 \Mpch$ and $N=1536^3$ particles, yielding a mass resolution $\mp=3\cdot 10^{11} h^{-1} M_\odot$. We refer to this set as L2400. All simulations where initialised with 2LPT at an initial redshift $z_i=49$.

The halo finding procedure is the same as the one described in L15. Halos are identified using the spherical overdensity halo finder Amiga Halo Finder (AHF) \cite{Gill:2004, Knollmann:2009} with an overdensity threshold $200\rho_m$ for the halo definition ($\rho_m$ is the background matter density). We bin the mass range of halos in 11 tophat bins of width 0.2 in logarithmic scale centered from $\lg M =12.55$ to $\lg M=14.55$, where $\lg$ is the base 10 logarithm. We use the L500 set for results in the range $\lg M =12.55 \, - \,\lg M=12.95$ and the L2400 set for higher masses. Hence, the lowest mass bin is centered on halos with around 51 particles for the L500 set and 47 for the L2400 set. We refer the reader to L15 for more details and justification of our choices.

\subsection{Measuring the bias parameters}
\label{sec:measurements}

In order to estimate the bias parameters from \refeq{biasestimator}, we need to measure  the linear and nonlinear matter density fields, as well as the halo density field. For the former, we generate the density field from the Zel'dovich displacement corresponding to the initial conditions of the given simulation at $z=99$, and linearly scale it with the growth factor $D$ to the final redshift. The nonlinear density field and halo fields are obtained from the simulation output and halo catalogs at the final redshift. We compute all of these fields on a grid of size $N_g=512$ for the L500 set and $N_g=764$ for the L2400 set. We can then construct all the relevant operators $O$ by going back and forth from real to Fourier space. Spatial derivatives and nonlocal operators such as $1/\nabla^2$ are applied in Fourier space, while products are taken in real space. The operators are constructed from the linearly evolved initial density field smoothed with a Gaussian filter on the scale $R=15\Mpch$. 

We then compute all the needed power spectra of the operators and halo fields, and evaluate the matrix $\v{M}$ and vector $\v{H}$ for multiple $k$ bins. Finally  $\v{C}(k)$ is obtained from \refeq{biasestimatork}. Each of the bias combinations entering $\v{C}(k)$ is expected to asymptote to a constant at low $k$. In order to maximize the signal-to-noise ratio while ensuring robust results, we perform a quadratic fit of the form $C_O(k)=c_O+A_O k^2$ up to $k_{\rm max}=0.18\Mpch$ for each of the components $C_O$ of $\v{C}$. The constant coefficients $c_O$ are the desired combinations of bias parameters given in \refeq{C}, while the coefficients $A_O$ are left free to absorb higher-order corrections to the correlators. We verified the robustness of our results under changes of the smoothing scale as well as $k_{\rm max}$. These consistency tests are presented in \refapp{convergence}.

In order to obtain an optimal fit of $C_O(k)$, we weight points at each $k$ value by their inverse variance. For the L500 set, we obtain the latter by a bootstrap procedure using $12'000$ random resamples of 48 realisations. In each $k$ bin and for each parameter $C_O(k)$ we compute the mean of each resample. We then compute the mean and standard deviation of the mean of the means distribution. These are the points and error used for the fit. The mean and error bars of our measurements are obtained by a second bootstrap, this time over the fit of each bias parameter, in a similar fashion as outlined in L15. We again create $12'000$ random resamples of 48 realisations each and compute the mean of means and its standard deviation, which are the final results we present. Notice that this procedure yields the marginalized error bars (see \refapp{covariance}). We cannot build robust error bars for the L2400 set in the same way since it contains only two realisations. Hence we rescale the error bars obtained with the L500 set at each $k$, $[\sigma(C_O(k))]_{\rm L500}$, by the total volume of each set. That is 
\be
[\sigma(C_O(k))]_{\rm L2400}=\sqrt{\frac{V_{\rm L500}}{V_{\rm L2400}}}\,[\sigma(C_O(k))]_{\rm L500}\, , 
\label{eq:errorscaling}
\ee
where $V_{\rm L500}=48\cdot500^3 (\Mpch)^3$ and $V_{\rm L500}=2\cdot2400^3 (\Mpch)^3$. Since the statistical error bars on the final parameters of interest $c_O$ are expected to scale the same way with volume, we perform the same rescaling for $\sigma(c_O)$. Note that we use the same smoothing scale and wavenumber bins for both L2400 and L500. Hence this rescaling via the simulation volume is expected to be accurate.

The procedure to obtain the bias parameters from the squared-field method is exactly analogous, except that we do not subtract $b_1\dm$ from the halo density field. This allows us to obtain a measurement for $b_1$ [\refeq{Csq}] and hence the complete set of bias parameters up to third order. 

Finally, we are interested in individual bias parameters entering \refeq{biasexp} at third order rather than the combinations in \refeq{C} (and \refeq{Csq} for squared-field). For this we use the precise measurements of the LIMD bias parameters $b_1^\text{SU}, b_2^\text{SU}$ presented in L15 (obtained using separate universe (SU) simulations \cite{Wagner:2014}), which were computed for the same cosmology and halo finder parameters as the present work. These measurements have comparable or smaller statistical errors than those obtained on the same parameters from the cubed- and squared-field methods used here, and are expected to be more robust to systematic errors as well. Specifically, in the squared-field case, $b_2$ and $b_{K^2}$ are obtained by subtracting $b_1^{\rm SU}$ from $c_{\d^2}$ and $c_{K^2}$ respectively. For the cubed-field case, $b_3$ is obtained by subtracting $b_2^\text{SU}$ from $c_{\d^3}$ whilst $b_{\d K^2}, \, b_{K^3}$, and $\btd$ are obtained by subtracting $b_{K^2}$ obtained from squared fields from $c_{\d K^2}$, $c_{K^3}$, and $c_{\otd}$ respectively (as well as $b_2^\text{SU}$ from L15 in the case of $b_{\d K^2}$). The error on the individual parameters $b_O$ is obtained by Gaussian error propagation from the error on $c_O$ and on $b_n^{\rm SU}$.

\section{Previous measurements and predictions}
\label{sec:predictions} 

In this section we review previous measurements as well as model predictions for some of the bias parameters which we measure. We will not focus on results for the LIMD bias parameters $b_1$, $b_2$, and $b_3$ in the main text (since these have already been extensively studied in the past, e.g \cite{Matarrese:1997sk, Angulo:2007, guo/jing:2009, tinker/etal:2010,pollack/smith/porciani:2012, Saito:2014qha,Lazeyras:2015, Li:2015, Baldauf:2015,hoffmann/etal:2015,Hoffmann:2016omy}), and hence we do not present previous measurements or model predictions for these here. We present our results for these parameters in the form of consistency checks in \refapp{check} by comparing them with the results of L15 which were obtained for the same cosmology. 

\subsection{Lagrangian local-in-matter-density (LLIMD) prediction}
\label{sec:limd}
The so-called Lagrangian local-in-matter-density (Lagrangian LIMD or simply LLIMD hereafter) model provides predictions for all bias parameters given a set of Lagrangian LIMD parameters $b_n^L \equiv n! b_{\d^n}^L$. We briefly recap this ansatz here but refer the reader to sections 2.2-2.4 of \cite{biasreview} for more details. This model assumes that halos formed instantaneously at some high redshift, and that their formation is exclusively governed by the matter density field in their neighborhood, i.e 
\be
\d_h^L(\vq)=b^L_1\d(\vq)+\frac12b^L_2\d^2(\vq)+\frac16b^L_3\d^3(\vq)+\dots \, ,
\label{eq:localbiasexp}
\ee
where $\vq$ is the Lagrangian position. That is, any influence of the tidal field on the proto-halo locations in Lagrangian space is assumed to be negligible. Using the fact that halos and matter comove on large scales (as required by the equivalence principle), we can then solve the continuity equation for both halos and matter with the same peculiar velocity divergence $\theta=\partial_iv^i$, yielding
\be
\frac{1}{1+\d_h}{\rm D}_\tau\d_h=\frac{1}{1+\d}{\rm D}_\tau\d \, ,
\ee
with ${\rm D}_\tau=\partial_\tau+v^i\partial_i$ denoting the convective derivative. We can write the solution of this equation in terms of the matter density field at the initial and final times and the halo density field at initial time, where final and initial positions are related through the trajectory of the matter fluid. By then inserting our prescription for the initial halo field \refeq{localbiasexp} in this solution, we obtain an expression for the Eulerian halo field as a function of Eulerian operators multiplied by Eulerian bias parameters expressed in terms of the Lagrangian ones. Interestingly, gravitational evolution sources terms that involve the tidal field $K_{ij}$, showing that the Lagrangian LIMD ansatz is inconsistent with the Eulerian LIMD picture \cite{catelan/etal:1998,catelan/porciani/kamionkowski:2000}. The solution at second order reads \cite{sheth/chan/scoccimarro:2012} 
\be 
\d_h^{(1+2)}=(1+b_1^L)\d^{(1+2)}+\left(\frac{4}{21}b_1^L+\frac12b_2^L\right)\d^2-\frac27b_1^L K^2.
\ee
We identify the term multiplying $\d^{(1+2)}$ as $b_1$, the one multiplying $\d^2/2$ as $b_2$, and the one multiplying $K^2$ as $b_{K^2}$. The same solution at third order gives a prediction for $b_{\d K^2}, b_{K^3}, \btd$. The final results for the Lagrangian LIMD prediction are \cite{MSZ,biasreview}
\ba
b_{K^2}^{\rm LLIMD} & = -\frac27\left(b_1-1\right),\\
\btd^{\rm LLIMD}+\frac{5}{2}b_{K^2}^{\rm LLIMD} & = -\frac16\left(b_1-1\right),\\
\btd^{\rm LLIMD} & = \frac{23}{42}\left(b_1-1\right), \\
b_{K^3}^{\rm LLIMD} & = \frac{22}{63}\left(b_1-1\right), \\
b_{\d K^2}^{\rm LLIMD} &= \frac{11}{49}\left(b_1-1\right)-\frac27\left[b_2-\frac{8}{21}\left(b_1-1\right)\right]\,.
\label{eq:LIMDbias}
\ea
For the numerical evaluation shown later, we use the best fit of L15 (their Eq.(5.2)) for the relation $b_2(b_1)$ in the last equation.

\subsection{Previous measurements}
\label{sec:prev}
As explained in \refsec{intro}, there are numerous previous measurements for $b_{K^2}$ from diverse techniques such as fit to the halo bispectrum or Lagrangian moments-based measurements \cite{chan/scoccimarro/sheth:2012,baldauf/etal:2012,Saito:2014qha,angulo/etal, sheth/chan/scoccimarro:2012, bel/hoffmann/gaztanaga:15,castorina/paranjape/etal:2016, Modi:2016dah}. We will compare our results for $b_{K^2}$  with the best fit of \cite{Modi:2016dah}. They used various Fourier space as well as real space methods to estimate the linear and quadratic Lagrangian bias parameters from numerical simulations. By then evolving these in time in the same fashion as what we presented for the LLIMD model in the previous section, they were able to give prediction for relations between the Eulerian biases. The one of interest for us is their Eq. (22) relating $b_{K^2}$ to $b_1$. Since they found a nonzero Lagrangian $b_{K^2}^L$ their results trivially disagree with the Lagrangian LIMD prediction.

Finally, Ref.~\cite{Saito:2014qha} used a joint fit to the tree-level halo-matter-matter bispectrum and the 1-loop halo-matter power spectrum to measure both $b_{K^2}$ and the combination
\be
b_{3\rm nl} \equiv -\frac{64}{105}\left(\btd+\frac52 b_{K^2}\right)\,,
\label{eq:b3nl}
\ee
and provided tabulated values for these (see their table 1). Expressions for the tree-level bispectrum and 1-loop halo-matter power spectrum can be found in Sec.~4.1.1 and Sec.~4.1.4 of \cite{biasreview}, respectively.

The tree-level halo-matter-matter bispectrum depends on the parameters $b_1$, $b_2$ and $b_{K^2}$, while the 1-loop power spectrum further depends on $\btd$ (more precisely the combination $b_{K^2}+2/5\btd$) as well as the higher-derivative bias $\blapl$, as explained in \refsec{intro}. In particular, the contributions involving the latter two bias parameters are highly degenerate in shape. To break this degeneracy, Ref.~\cite{Saito:2014qha} used a joint fit of the power spectrum and bispectrum, and, crucially, set $\blapl=0$. This last assumption is not expected to be correct and hence might bias their measurements of $b_{3\rm nl}$.  
Further, they show clearly that adding the dependence of the 1-loop power spectrum on $b_{K^2}$ does not change the best-fit value for this quantity significantly. This means that, although they perform a joint fit, $b_{K^2}$ is effectively determined by the tree-level bispectrum while the combination $b_{K^2}+2/5\btd$ is obtained from the 1-loop power spectrum. Since $\blapl$ only enters the latter, we expect that their assumption of $\blapl=0$ mainly affects their results for $b_{3\rm nl}$.  

\section{Results and discussion}
\label{sec:results}

In this section, we present our results for the four tidal bias parameters that our method allows us to measure. We show measurements at redshifts $0, \, 0.5$ and $1$. We present results for $b_1$, $b_2$ and $b_3$ at $z=0.0$ and compare them to the ones of L15 in \refapp{check} as checks of our method. 

\subsection{$b_{K^2}$}
\label{sec:bk2btd} 

\begin{figure}
\centering
\includegraphics[scale=0.4]{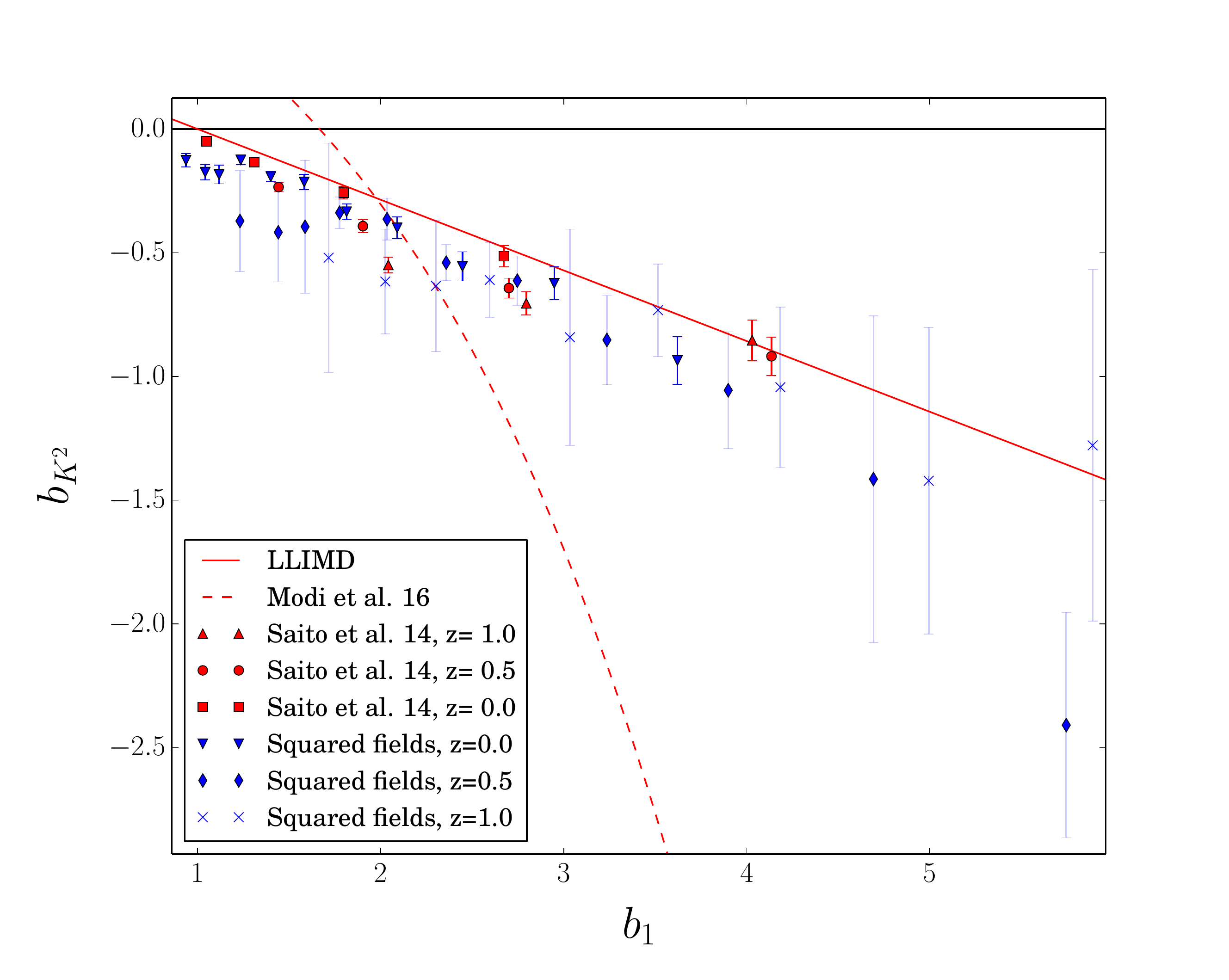}
\caption{$b_{K^2}$ as a function of $b_1$. The blue symbols present our results for this bias parameter at different redshifts while the red ones are the measurements from \cite{Saito:2014qha}. The dashed and solid red lines present the best fit from \cite{Modi:2016dah}, and the LLIMD prediction respectively. Our results are in excellent agreement with the ones from \cite{Saito:2014qha} and show that the relation between these two parameters is approximately linear, as predicted by the LLIMD ansatz, with a slight systematic shift towards more negative values. The fitting relation based on the measurements from \cite{Modi:2016dah} is however in strong disagreement.   
See text for more details.}
\label{fig:bk2}
\end{figure}

We start with $b_{K^2}$, for which previous measurements exist. \refFig{bk2} presents the results for $b_{K^2}$ obtained from the squared-field correlators as a function of $b_1$. We show the comparison with the results from cubed-field correlators as well as convergence tests in \refapp{convergence}. This parameter is negative for all masses probed by our simulations, which reflects the anti-correlation between the tidal field and halo field. Comparing our results to the LLIMD prediction, we observe a slight systematic shift of $b_{K^2}$ towards more negative values. This indicates a roughly mass-independent negative value of the Lagrangian tidal bias $b_{K^2}^L$. This result thus shows that the tidal field already has to be taken into account in Lagrangian space, in contradiction with the LLIMD assumption. Nevertheless, this is expected physically, at least for halos with $b_1^L>0$, since the tidal field elongates proto-halos in a given direction making the collapse to a halo more difficult. 

We further compare our results to the best fit of \cite{Modi:2016dah} as well as with the measurements obtained from the tree-level bispectrum in \cite{Saito:2014qha}. Our results are in excellent agreement with the ones from \cite{Saito:2014qha}, especially given that their results were obtained for a different simulation cosmology and a Friends-of-Friends halo finder. Comparing the error bars between the two sets of simulations shows that our method is competitive with theirs. Notice however that we use a total simulation volume which is roughly $2/3$ of theirs and a maximum $k$ for the fit of 0.18 Mpc$/h$ compared to their 0.125 Mpc$/h$ for the power spectrum, and 0.065 Mpc$/h$ for the bispectrum; on the other hand, we use the cross-correlation of the halo field with the linearly evolved matter density field. The best fit from \cite{Modi:2016dah} is however in strong disagreement with our results. Note that their measurement is based on measuring moments of halo counts and the density and tidal field in subvolumes of the simulation box. This method is fairly different from the squared-field and bispectrum methods, which are both based on the  large-scale halo-matter-matter three-point function. This disagreement clearly warrants further investigation. Finally, we find good agreement for this parameter with the results of \cite{Abidi:2018eyd}.

\begin{figure}
\centering
\includegraphics[scale=0.4]{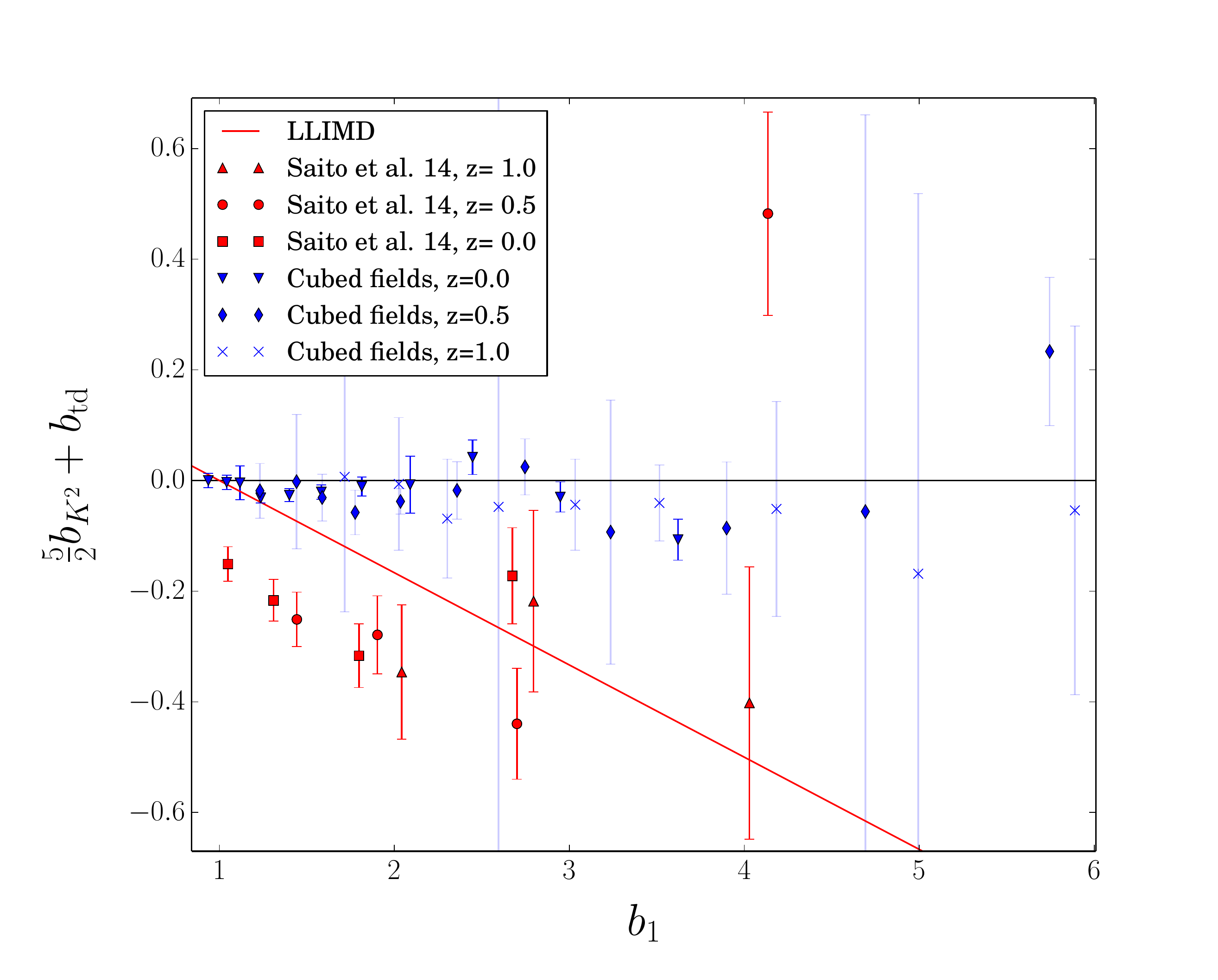}
\caption{$\btd+\frac52 b_{K^2}$ as a function of $b_1$. The blue symbols show results obtained from cubed-field correlators at various redshifts, while the red ones show results from \cite{Saito:2014qha}, and the line indicates the LLIMD prediction. The agreement between the two sets of measurements is much worse than for $b_{K^2}$, for reasons that we explain in the text.}
\label{fig:bk2btd}
\end{figure}

\subsection{$\btd$}
\label{sec:btd}

We next present results for the combination $\btd+\frac52 b_{K^2}$ as a function of $b_1$ in \reffig{bk2btd}. These are obtained from correlators of cubed fields as outlined in \refsec{biasest} and \refsec{measurements}.  Ref. \cite{Saito:2014qha} also presented measurements for this combination, via their $b_{3\rm nl}$ defined in \refeq{b3nl}. As \reffig{bk2btd} shows, we find this combination of bias parameters to be consistent with zero. The agreement between our measurements and both the results of \cite{Saito:2014qha} and the LLIMD prediction is less good than in the case of $b_{K^2}$. Notice however that the disagreement with \cite{Saito:2014qha} could be explained by the fact that their results on this combination of bias parameters comes from the 1-loop power spectrum under the assumption that the higher-derivative bias vanishes, $\blapl=0$. Hence, the disagreement between our measurements and the results of \cite{Saito:2014qha} could indicate that $\blapl$ is in fact nonzero. Nevertheless, it would be interesting to investigate possible explanations for the fact that the result of \cite{Saito:2014qha} is close to Lagrangian LIMD. Moreover, given the substantial evidence for a departure of $b_{K^2}$ from LLIMD, one might expect a similar departure for $\btd$ as well. Clearly, however, our results indicate a stronger deviation from LLIMD in this linear combination of bias parameters than that seen in $b_{K^2}$. Finally, our results for this combination of parameter is consistent with those of \cite{Abidi:2018eyd} who found it to be consistent both with zero and with the LLIMD prediction (see their Figure 12).

\begin{figure}
\centering
\includegraphics[scale=0.4]{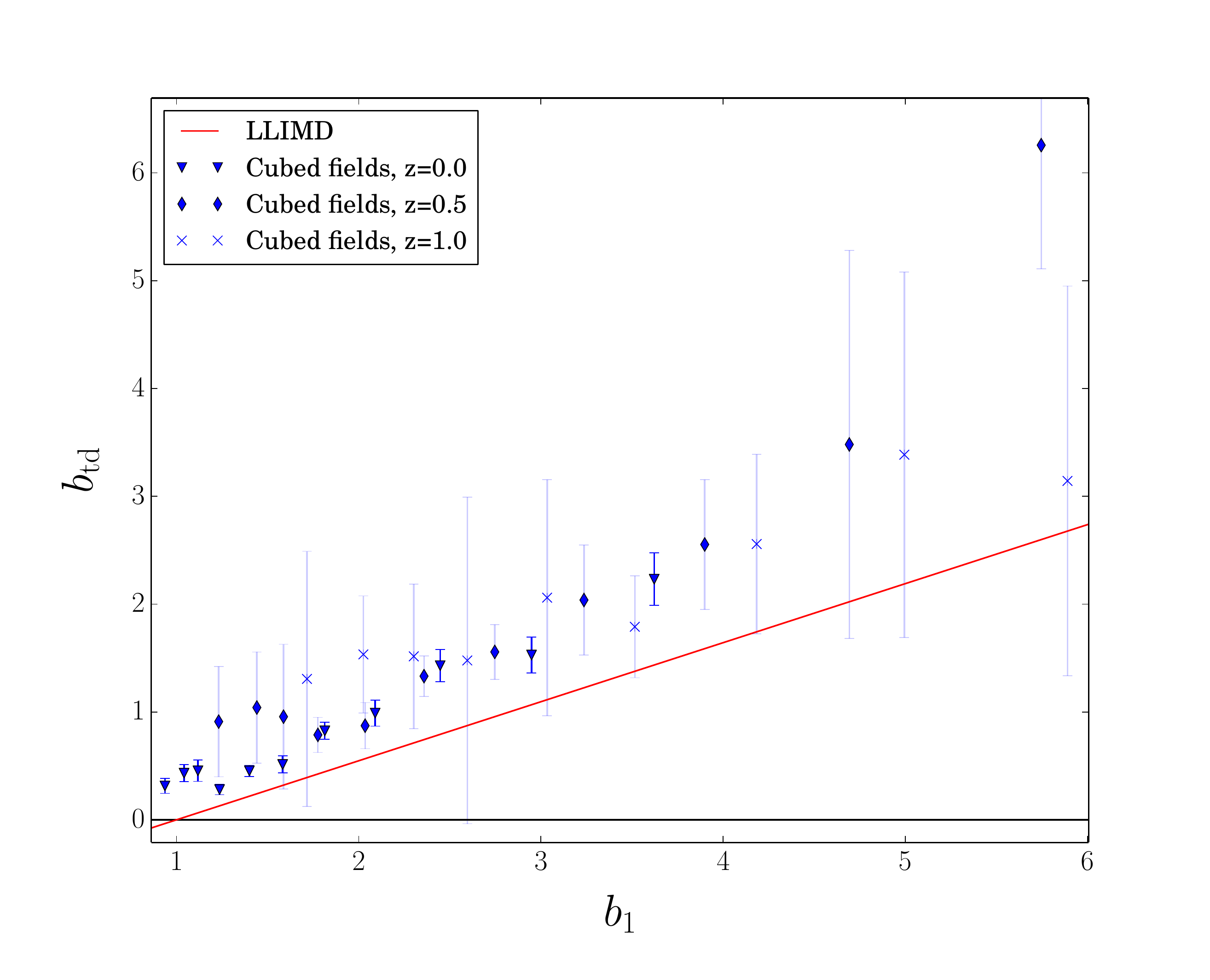}
\caption{$\btd$ as a function of $b_1$. The color coding is the same as in \reffig{bk2btd}.}
\label{fig:btd}
\end{figure}

We can further obtain results for $\btd$ alone by subtracting the results for $b_{K^2}$ from the ones presented in \reffig{bk2btd}. This is presented in \reffig{btd} as a function of $b_1$. Notice that this is the first time that results for this bias only have been obtained. We again have a clear detection of this parameter which is consistent with being positive at all halo masses, and is slightly larger than the Lagrangian LIMD prediction. 

\subsection{$b_{\d K^2}$ and $b_{K^3}$}
\label{sec:bk3bk2d}

\begin{figure}
\centering
\includegraphics[scale=0.4]{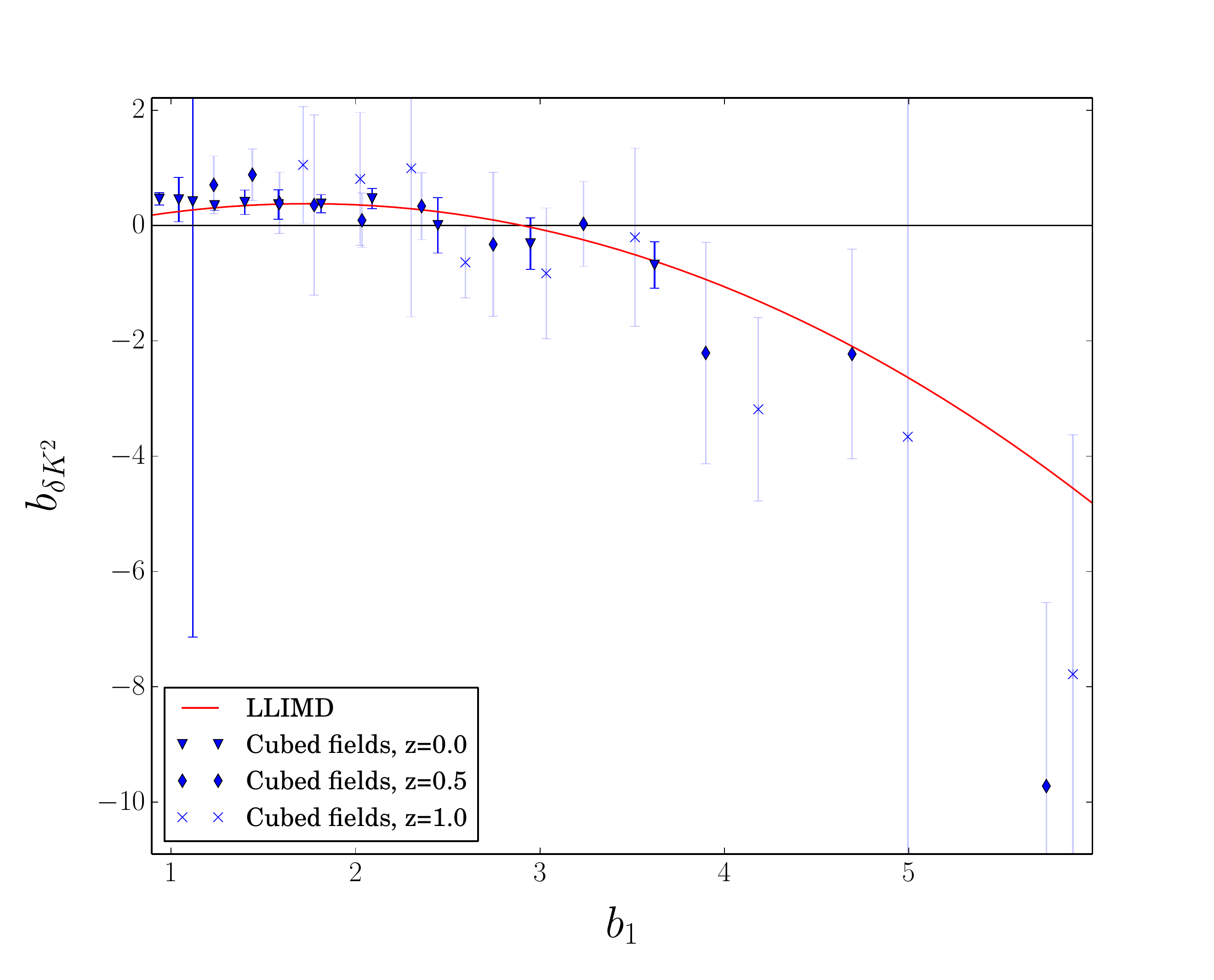}
\caption{$b_{\d K^2}$ as a function of $b_1$. The color coding is the same as in \reffig{bk2btd}.}
\label{fig:bdk2}
\end{figure}

\begin{figure}
\centering
\includegraphics[scale=0.4]{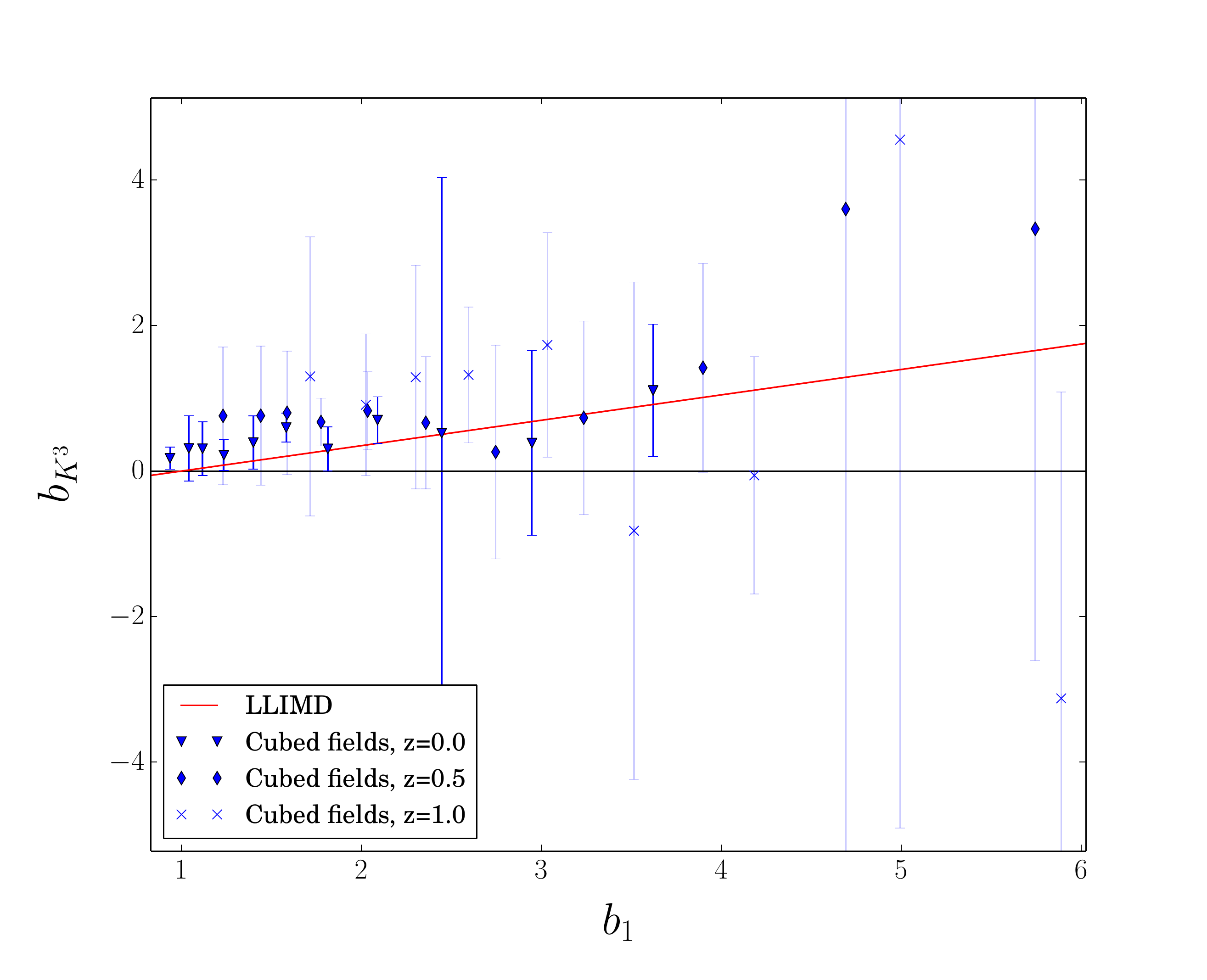}
\caption{$b_{K^3}$ as a function of $b_1$. The color coding is the same as in \reffig{bk2btd}.}
\label{fig:bk3}
\end{figure}

We now turn to the two remaining bias parameters, namely $b_{K^3}$ and $b_{\d K^2}$ for which we present the first measurements to date. These were obtained from the cubed-field method, and are shown in \reffigs{bdk2}{bk3} as a function of $b_1$. Again, we see a clear detection of both of these bias parameters, especially for $b_{\d K^2}$. While these bias parameters do not enter the 1-loop halo power spectrum and are thus less relevant for the large-scale statistics of halos than $b_{K^2}$ and $\btd$, these results can inform physical models of halo formation. We find good agreement within errors with the LLIMD prediction for both bias parameters, which is also in agreement with the results of \cite{Abidi:2018eyd} for two related bias parameters ($b_{\mathcal{G}_3}$ and $b_{\d\mathcal{G}_2}$).

\section{Conclusions}
\label{sec:concl} 

Using 2-point correlators of quadratic and cubic operators constructed out of the linear density and tidal fields, we have presented new measurements of the complete set of local bias parameters entering the bias expansion up to third order. Our method not only allows one to measure a number of bias parameters efficiently, but it is also competitive with other methods in terms of constraining power (as can be seen by comparing the size of the error bars in \reffig{bk2}). We present the first measurements to date for $\btd$, $b_{K^3}$ and $b_{\d K^2}$. These results are very encouraging and will hopefully be compared with independent measurements in the near future.

We have compared our measurements for the LIMD bias parameters $b_n$ with those of L15 in \refapp{check} and found good agreement, validating our method. Furthermore, the convergence tests presented in \refapp{convergence} confirm the stability of our results under the change of the parameters entering our analysis. Regarding the bias parameters involving the tidal field, our main findings are:
\begin{itemize}
\item We find excellent agreement between our results and those of \cite{Saito:2014qha} for  $b_{K^2}$. These results confirm a linear relation between this parameter and the linear LIMD bias $b_1$, as predicted by Lagrangian LIMD. We however find a small negative constant offset between our measurements and the LLIMD prediction, implying an approximately mass-independent Lagrangian tidal bias $b_{K^2}^L < 0$, consistent with the findings of \cite{Saito:2014qha}.
\item The moments-based results for $b_{K^2}$ presented in \cite{Modi:2016dah} do not agree with our results, nor with \cite{Saito:2014qha}. The source of disagreement is unclear at this point and clearly warrants further investigation.
\item We find the combination $\btd+5/2b_{K^2}$ to be consistent with zero. The agreement between our results and those of \cite{Saito:2014qha} is much worse than for $b_{K^2}$. However, as explained in \refsec{prev}, this could come from the fact that they set $\blapl=0$. Our results would then indicate a nonzero $\blapl$ (see also \cite{angulo/etal,Fujita:2016,Abidi:2018eyd}). 
\item We also obtain a clear detection for $\btd$ being nonzero. Given the degeneracy with $\blapl$ in the halo power spectrum, the result presented here is the first direct measurement of this bias parameter in the literature. As in the case of $b_{K^2}$ we find good agreement between our results and the LLIMD prediction with a small systematic shift indicating a nonzero Lagrangian bias $b_{\rm td}^L > 0$.
\item Finally, we also obtain the first measurements of $b_{\d K^2}$ and $b_{K^3}$. The agreement between our results and the LLIMD prediction is quite good for both parameters.
\end{itemize}

Our efficient trispectrum estimator has thus opened substantial new territory in the field of halo bias. For example, our results will finally allow for a robust determination of the higher-derivative bias parameter $\blapl$. This parameter is of great interest since it involves a new scale, the scale on which halo formation occurs. With these results in hand, it will also become possible  to independently determine the reach of perturbation theory predictions for the halo power spectrum, since all free parameters have been fixed through the bispectrum and trispectrum. Further, more detailed studies of the stochasticity inherent to the halo formation process will become possible. Finally, our results on the bias parameters can inform analytical models of halo formation, such as excursion-set, peaks, and peak-patch approaches. These are only a few examples of the future implications of the robust measurements of higher-order bias presented here. 

\acknowledgments{We thank Christian Wagner for putting the L2400 set of simulations at our disposal, and Muntazir Abidi and Tobias Baldauf for helpful discussions after our first arXiv submission. TL thanks Alexandre Barreira and Aniket Agrawal for insightful discussions during the project. FS acknowledges support from the Marie Curie Career Integration Grant  (FP7-PEOPLE-2013-CIG) ``FundPhysicsAndLSS,'' and Starting Grant (ERC-2015-STG 678652) ``GrInflaGal'' from the European Research Council. 
}

\appendix

\section{Bias expansion to $3^{\rm rd}$ order}
\label{app:biasexp} 

In this appendix we present a short derivation of the complete bias expansion up to third order. This will show which particular combinations of bias parameters are returned by our bias estimators, as presented in the main text. For sake of clarity we drop the time and position arguments of the fields and we denote the linear fields $\d^{(1)}$ and $K_{ij}^{(1)}$ simply by $\d$ and $K_{ij}$, respectively. 

We start from \refeq{biasexp}. The complete set of operators up to third order is
\be 
O \in \left\{\dm, \lapl\dm, \dm^2, \Km^2, \dm^3, \dm \Km^2, \Km^3,\otd\right\},
\label{eq:operatorset}
\ee
where $\dm$ denotes the nonlinear (evolved) matter density field, while $\Km_{ij} = \mathcal{D}_{ij} \dm$ [cf.~\refeq{kij}] denotes the nonlinear tidal field, to be distinguished from the corresponding linearly evolved quantities $\d$ and $K_{ij}$, respectively. The definition of $\otd$ is given in \refeq{otd}. 
Note that $\otd$ is a cubic-order operator, and hence the leading-order result \refeq{otd} is sufficient for our purposes. In \refeq{operatorset}, we have included the leading higher-derivative contribution $\lapl\dm$, which leads to contributions to halo statistics that are of similar order as those from the cubic bias parameters. However, as we show in \refapp{HO}, these contributions are suppressed by $k^2$ compared to the other correlators,
and are thus absorbed by our scale-dependent correction $A_O k^2$. We have formally checked this by repeating our analysis including $\lapl\d$ in the set of operators \refeq{operatorset} and found the results to be completely consistent with the ones presented in \refsec{results}, albeit with lower signal to noise. Furthermore, we did not find a significant detection of a nonzero $\blapl$. We thus drop $\lapl\dm$ from the list in the following. 

The bias expansion is then 
\be
\d_h=b_1\dm+\frac12 b_2 \dm^2+b_{K^2} \Km^2 +\frac16 b_3 \dm^3 + b_{\d K^2}\d_m \Km^2 + b_{K^3} \Km^3 + \btd\otd \, .
\label{eq:biasexp3rdorder}
\ee
We are only interested in going to cubic order in \textit{linear} fields. Hence we expand $\dm$ as $\dm=\d+\d^{(2)}+\d^{(3)}+\dots$ (and hence also $\Km_{ij}=K_{ij}+K^{(2)}_{ij}+K^{(3)}_{ij}$). Inserting this into \refeq{biasexp3rdorder} yields   
\ba
\d_h&= \, b_1\left(\d+\d^{(2)}+\d^{(3)}\right)+\frac12 b_2 \left(\d^2+2\d\d^{(2)}\right)+b_{K^2} \left(K^2+2K^{(2)}_{ij}K^{ij}\right) +\frac16 b_3 \d^3 \nonumber \\ &+ b_{\d K^2}\d K^2 + b_{K^3} K^3 + \btd\otd \, , 
\label{eq:biasexp3rdorderlin}
\ea
where $\d^{(2)}$ and $K_{ij}^{(2)}$ are given by (see App.~B--C in \cite{biasreview})
\ba
\d^{(2)}&= \frac{17}{21} \d^2 + \frac27 (K_{ij})^2 
-s^i\partial_i\d \,,\\
K_{ij}^{(2)}&= \, \frac{10}{21} \Del_{ij} \left[ \d^2 - \frac32 K^2 \right]
+ K_{ik} K^k_{\  j} - \frac13 \d_{ij} K^2 + \frac23 \delta K_{ij} -  s^k \partial_k K_{ij}\,, \\
\Rightarrow\quad K^{ij} K^{(2)}_{ij} &= \, \frac{5}{4} \otd
+ K^3 + \frac23 \d K^2 - \frac12 s^k \partial_k K^2 ,
\label{eq:d2d3}
\ea
where $\v{s}_i=-(\partial_i/\lapl)\d$ is the displacement field. Note that the third-order matter density field involves several additional displacement operators. However, since we only consider the combination $\d_h-b_1\dm$ for the cubic correlators, these terms are consistently subtracted out, and we in fact do not need the expression for $\d^{(3)}$ here. We now insert these expressions into \refeq{biasexp3rdorderlin}, and reorganise the terms by operators to obtain the quantities of interest, $\d_h^{(2)}$ and $\left(\d_h-b_1\dm\right)^{(3)}$:
\ba
\d_h^{(2)}= & \, \left(\frac{17}{21}b_1+\frac12 b_2\right)\d^2 + \left(\frac27 b_1+b_{K^2}\right)K^2
-b_1 s^i \partial_i \d ,\\
\left(\d_h-b_1\dm\right)^{(3)} = & \, \left(\frac{17}{21}b_2+\frac16 b_3\right)\d^3+\left(\frac52 b_{K^2}+\btd\right)\otd^{(3)}+\left(b_{K^3}+2b_{K^2}\right) K^3 \nonumber \\
&+\left(b_{\d K^2}+\frac43 b_{K^2}+\frac27 b_2\right) \d K^2
-b_{K^2} s^i \partial_i K^2-\frac12 b_2 s^i \partial_i \d^2 .
\label{eq:biasexpfinal}
\ea
Hence the set of bias combinations one obtains from quadratic fields (without subtraction of $b_1 \dm$) is given by 
\be
\v{c}^{(2)} = \left(\begin{array}{c}
  b_2/2 + (17/21) b_1  \\
  b_{K^2} + (2/7) b_1\\
  - b_1 
  \end{array}
  \right)\,,
\ee
whilst the cubic ones (with subtraction of $b_1\dm$) yield
\be
\v{c}^{(3)} = \left(\begin{array}{c}
  b_3/6 + (17/21) b_2 \\
  b_{\d K^2} + (2/7) b_2 + (4/3) b_{K^2} \\
  b_{K^3} + 2 b_{K^2}  \\
  b_{\rm td} + (5/2) b_{K^2} \\
  - b_2/2 \\
  - b_{K^2}
  \end{array}
  \right)\,.
\ee

\section{Renormalization of operators}
\label{app:subtraction}

The cubic bias parameters appear in the tree-level trispectrum, which is the connected part of the 4-point function. When measuring the trispectrum through the cubed-field method, we need to ensure that the disconnected part of the 4-point function does not contribute. This can be ensured by including the leading counter-terms to the operators constructed from cubic combinations of the density field. The leading counter-terms are sufficient, since the cubic operators are constructed from the linear density field $\d(\vx)$.

Consider one of the cubic operators $O$ used in the cubed-field estimator, for a smoothing scale $R$, which we will not write explicitly for clarity, and assume that we can construct this operator such that
\be
\< [O](\vk) \d_{R'}(\vk') \> = 0\,,
\label{eq:subcond}
\ee
where $R'$ can be different from $R$, and the brackets around $O$ indicate that this is the renormalized operator. Note that this corresponds to one of the renormalization conditions derived in \cite{assassi/etal}, since the leading-order cross-correlation of a cubic operator with the density field is zero. 
It is then clear that the correlator $ \< [O](\vk) \d_h(\vk') \>' $, appearing on the left-hand side of \refeq{dhO3}, only contains connected trispectrum contributions. In particular, no linear-order higher-derivative bias terms contribute, which we have not removed by subtracting $b_1 \d_m$ from $\d_h$ in \refeq{dhO3}, since they simply correspond to powers of $k^2$ multiplying \refeq{subcond}.

Next, consider the cross-correlation of $[O](\vk)$ with one of the other cubic operators, $[O']$, as on the right-hand side of \refeq{dhO3} [\refeq{MOO}]. We can write the operators in Fourier space as
\be
[O](\vk) = \int_{\vp_1,\vp_2,\vp_3} \!\!\!\!\!(2\pi)^3 \d_D(\vk-\vp_{123})  S_{[O]}(\vp_1,\vp_2,\vp_3) \d_R(\vp_1)\d_R(\vp_2)\d_R(\vp_3)\,,
\ee
where $S_{[O]}$ is a kernel which includes the counter-terms. We then obtain 
\ba
\< [O](\vk) [O'](\vk') \>' =\:& \int_{\vp_1,\vp_2,\vp_3} \!\!\!\!\!(2\pi)^3 \d_D(\vk-\vp_{123})  S_{[O]}(\vp_1,\vp_2,\vp_3) \vs
& \times \int_{\vp'_1,\vp'_2,\vp'_3} \!\!\!\!\!(2\pi)^3 \d_D(\vk-\vp'_{123})  S_{[O']}(\vp'_1,\vp'_2,\vp'_3) \vs
&  \times \< \d_{R}(\vp_1) \d_{R}(\vp_2) \d_{R}(\vp_3)
\d_{R'}(\vp'_1) \d_{R'}(\vp'_2) \d_{R'}(\vp'_3)
\>'\,.
\label{eq:OOfull}
\ea
Now, any of the contractions of the $\vp_i$ or $\vp'_j$ among themselves lead, in general, to factors of the form
\be
\int_{\vp} \< \d_R(\vp) \d_R(-\vp) \>' = \< \d_R^2(\vx) \>\,.
\ee
These are zero-lag contributions, which should always be absorbed by counter-terms in the renormalized bias expansion (see, e.g. \cite{mcdonald,PBSpaper,assassi/etal}). 
For any such contraction of the 6-point correlator in \refeq{OOfull} however, the resulting correlator is proportional to
\ba
& \int_{\vp_1,\vp_2,\vp_3} \!\!\!\!\!(2\pi)^3 \d_D(\vk-\vp_{123})  S_{[O]}(\vp_1,\vp_2,\vp_3) \< \d_{R}(\vp_1) \d_{R}(\vp_2) \d_{R}(\vp_3) \d_{R'}(\vk) \>' \times \< \d_{R'}^2 \> \vs
&= \<  O(\vk) \d_{R'}(\vk)\>' \times \< \d_{R'}^2 \> \,.
\ea
where we have assumed (without loss of generality) that two of the $p_i'$ are contracted. The integral over the kernel $S_{[O']}$ in \refeq{OOfull} simply yields a proportionality constant for these types of contractions. We see that the renormalization condition \refeq{subcond} ensures that all of the contractions involving zero-lag correlators $\< \d_R^2\>,\  \< \d_{R'}^2\>$ vanish.

Let us now consider how \refeq{subcond} can be satisfied. For this, it is simpler to use the real-space correlators. First, for $O = \d_R^3$, we have
\ba
\< \d_R^3(\vx) \d_{R'}(\vy) \> = 3 \< \d_R^2\> \< \d_R(\vx) \d_{R'}(\vy)\>\,.
\ea
It is clear that we can remove this unwanted contribution by replacing
\be
\d_R^3 \to [\d_R^3] = \d_R^3 - 3 \< \d_R^2\> \d_R\,.
\ee
Similarly, one easily finds
\ba
[\d_R (K_R)^2] &= \d_R (K_R)^2 - \< (K_R)^2\> \d_R \vs
\left[ (K_R)^3 \right]
&= (K_R)^3\,,
\ea
i.e., $(K_{ij})^3$ does not lead to disconnected contributions (recall that we are always constructing operators from the linear density field). Next, we have
\ba
\< (s_R^k \partial_k \,\d_R^2)(\vx) \d_{R'}(\vy) \> =\:& 2 \< (s_R^k\partial_k \d_R)(\vx) \> \< \d_R(\vx) \d_{R'}(\vy)\>\,,
\ea
leading to
\be
\left[s_R^k \partial_k\, \d_R^2\right] = s_R^k \partial_k\, \d_R^2 - 2 \< \d_R^2\> \d_R\,,
\ee
since $\< s_R^k\partial_k \d_R \> = \< \d_R^2\>$. Further,
\be
\left[s_R^k \partial_k\, (K_R)^2\right] = s_R^k \partial_k (K_R)^2 \,.
\ee

Finally, we turn to $\otd$. The cross-correlation with $\d_{R'}$ is given, in the notation of \cite{biasreview}, by
\ba
\< \otd(\vk) \d_{R'} (\vk') \> =\:& \frac25 f_{\text{NLO},R}(k) W_R(k) W_{R'}(k) \Plin(k) \vs
f_{\text{NLO},R}(k) =\:& 4 \int_{\vp} \left[\frac{[\vp\cdot(\vk-\vp)]^2}{p^2 |\vk-\vp|^2}-1\right] F_2(\vk,-\vp) |W_R(p)|^2 \Plin(p)\,.
\label{eq:otdcorr}
\ea
This is not zero, but since $f_\text{NLO}(k) \propto k^2$ on large scales, it is suppressed relative to the other zero-lag contributions. It is not simply removed by a subtraction of $\d_R$ in real space. However, it is not necessary to remove the contribution in \refeq{otdcorr}. First, for the cross-correlation of $\otd$ with the halo field, the only contribution relevant at this order comes from the linear-order $\d_h$, which we subtract in \refeq{dhO3}. Second, for the operator cross-correlations, no zero-lag contribution remains, since we subtract the corresponding terms from \emph{all other} cubic operators. 

These considerations finally lead to \refeq{subtr}.

\section{Higher-order corrections}
\label{app:HO}

In this appendix, we investigate higher-order contributions neglected in our analysis which could potentially bias the measurements of bias parameters. 
Since the cubic operators are constructed from the linearly evolved density field, there are no higher-order corrections to the operator cross-correlations. Thus, we only need to consider possible higher-order contributions to $\< \d_h(\vk) [O](\vk')\>'$.

First, let us consider higher-derivative operators that appear in the higher-order bias expansion of $\d_h(\vk)$. At linear order in perturbations, these have the form
\be
\d_h(\vk) \supset \left[\sum_{n=1}^\infty (-1)^n b_{\nabla^{2n}\delta} k^{2n}\right] \d(\vk)\,.
\ee
As explained in the text, we expect $b_{\nabla^{2n}\delta}$ to be of order $R_L^{2n}$, where $R_L$ is the Lagrangian radius of halos.
We see that \refeq{subcond} is sufficient to ensure that none of these contribute to our bias estimation at any order. We have formally checked this for the case of $\blapl \lapl \d$ by repeating our analysis including $\lapl\d$ in the set of operators \refeq{operatorset} and found the results to be completely consistent with the ones presented in \refsec{results}. At cubic order in perturbations, we further have higher-derivative contributions such as $\nabla^2 [O^{(3)}](\vx)$. These clearly correct \refeq{dhO3} by contributions that scale as $k^2$ times the operator correlators $\< [O](\vk) [O'](\vk')\>$, and are thus absorbed by the marginalization over the coefficient $A_O$ of the $k^2$ term in the bias estimate as a function of $k$ (\refsec{biasest}). There are other cubic higher-derivative contributions, which are not given by total derivatives on the cubic operators. However, they will still be suppressed by $R_L^2 k^2$ compared to the leading correlators.

We now turn to higher-order perturbative corrections. Since the cubic operators $O$ are constructed from the linear density field, the leading higher-order term involves the cross-correlation of $\d_h^{(5)}$ with $[O^{(3)}]$, which can be written as
\ba
\< \d_h^{(5)}(\vk) [O](\vk') \>'_\text{NLO} \:& = \left(\prod_{i=1}^3 \int_{\vp_i}\right) (2\pi)^3 \d_D(\vk'-\vp_{123}) \left(\prod_{j=1}^5 \int_{\vp'_j}\right) (2\pi)^3 \d_D(\vk-\vp'_{12345}) \vs
   \times & S_{[O]}(\vp_1,\vp_2,\vp_3) S_{\d_h^{(5)}}(\vp'_1,\cdots,\vp'_5)
\< \d_R(\vp_1)\d_R(\vp_2) \d_R(\vp_3) \d(\vp'_1)\cdots\d(\vp'_5)\>' .
\nonumber
\ea
The condition \refeq{subcond} on $[O](\vk)$ ensures that each of $\vp_1,\vp_2,\vp_3$ must be contracted with one of the $\vp'_j$. Assuming that the kernel $S_{\d_h^{(5)}}$ describing the fifth-order halo density field (which of course also contains many bias parameters) is fully symmetrized, this loop integral becomes
\ba
\< \d_h^{(5)}(\vk) [O](\vk') \>'_\text{NLO} =\:& 10\left(\prod_{i=1}^3 \int_{\vp_i}  \right) (2\pi)^3 \d_D(\vk'-\vp_{123})  S_{[O]}(\vp_1,\vp_2,\vp_3) \vs
& \times  \left[\int_{\vp} S_{\d_h^{(5)}}(\vp,-\vp,-\vp_1,-\vp_2,-\vp_3)
\Plin(p)\right] \prod_{i=1}^3 W_R(p_i) \Plin(p_i) \, .
\nonumber
\ea
We see that this is of similar form as the leading-order operator correlators, with the difference of an additional integral, or loop, in brackets in the second line (the integrals over $\vp_i$ are really just weighted combinations of different modes of the trispectrum, rather than loops). If this loop integral asymptotes to a constant in the limit of $p_i\to 0$, then it is a term that is absorbed by counter-terms to one of the cubic-order operators which are necessary to include at fifth order. Such loop contributions are thus irrelevant in the renormalized bias expansion.

The remaining, non-trivial loop contributions approximately scale as $(p_i/\knl)^2$ in the large-scale limit. This is analogous to the ``1--3''-contribution to the 1-loop halo power spectrum which involves the correlator in \refeq{otdcorr}. We thus expect that these higher-order contributions are also effectively absorbed by our marginalization of a correction $A_O k^2$ in our bias fit.

\section{Convergence tests}
\label{app:convergence}

In this appendix we present three convergence tests to verify the robustness of our results under the change of the parameters in our algorithm, i.e the smoothing scale $R$ and the maximum $k$ value used for the fit of $\v{C}(k)$. We also check the consistency of the results between the L500 and L2400 sets of simulations. We use the results for $b_2$ and $b_{K^2}$ at $z=0$ here, since we have results for both squared-field and cubed-field methods for these. This will allow us to also check the consistency between the two methods and explore possible systematic errors. The LIMD bias parameters $b_n$ will be considered in \refapp{check}.

\begin{figure}
\centering
\includegraphics[scale=0.28]{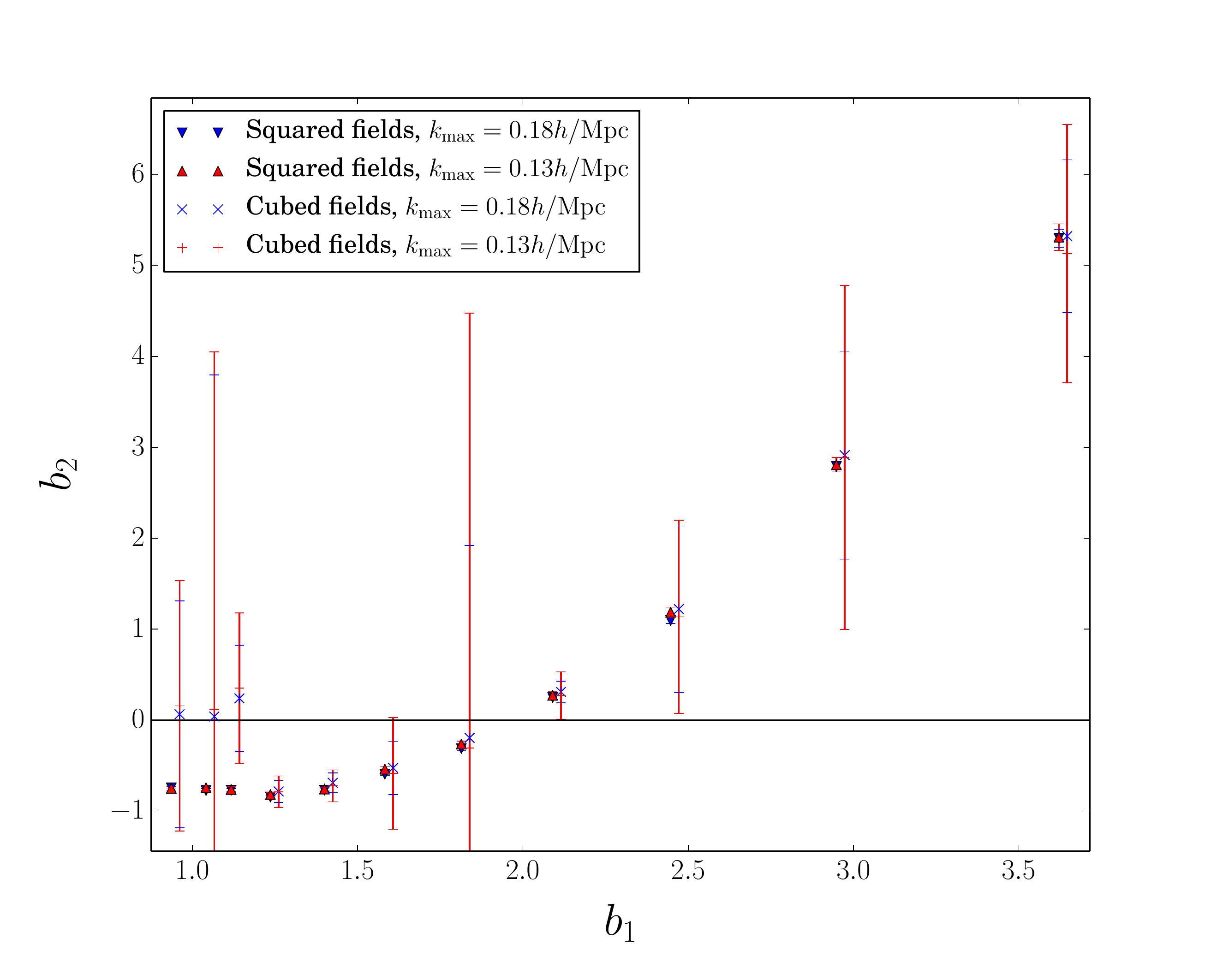}
\includegraphics[scale=0.28]{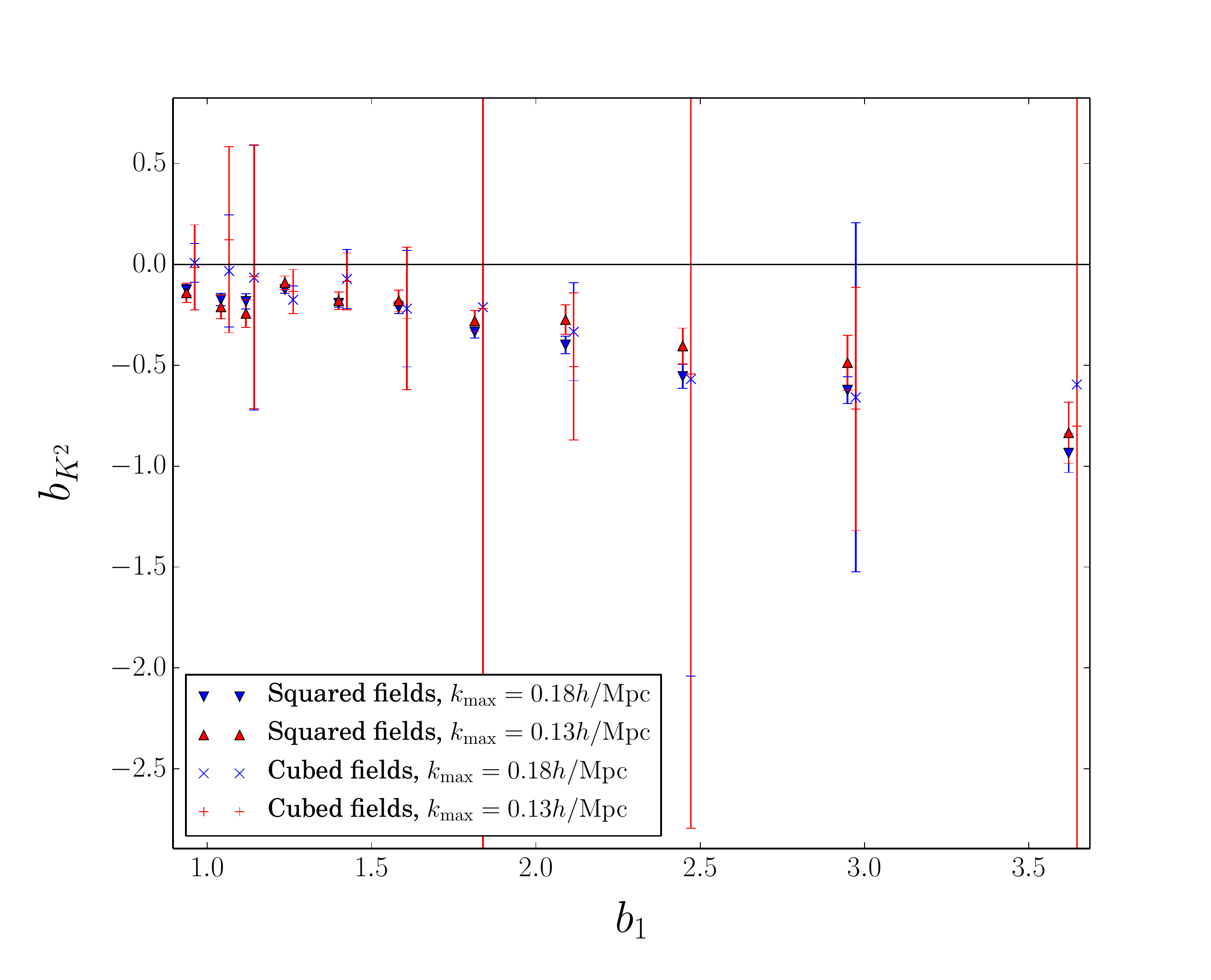}
\caption{$b_2$ (left) and $b_{K^2}$ (right) as a function of $b_1$ for two different maximum $k$ values for the fit. The triangles present results from the squared-field estimator while the crosses present those from the cubed-field estimator (slightly displaced horizontally for clarity), and the color coding indicates the $k_{\rm max}$ used. Results for the two different $k_{\rm max}$ within each method, as well between the two methods are largely self-consistent. The small discrepancy between the two methods for the three lowest values of $b_1$ comes from a lack of signal for the quadratic biases from the L500 set for cubed-fields, as can be seen in \reffig{Ldep}.}
\label{fig:kdep}
\end{figure}

We start by comparing results when fitting up to $k_{\rm max}=0.13$ and 0.18 $h/$Mpc in \reffig{kdep}. The results from squared and cubed fields are self-consistent for both bias parameters and prove the robustness of our measurements under a change in the fit range. As is expected, decreasing the maximum $k$ increases the error bars, but only mildly affects the mean. Furthermore, results  between the two methods are largely self-consistent. The small discrepancy between the two methods at low mass (for the three lowest values of $b_1$) comes from a lack of signal for the quadratic parameters from the L500 set for cubed-fields, as can be seen in \reffig{Ldep}. As can be seen in \reffig{cova}, $b_2$ correlates strongly with $b_3$ and $b_{K^2}$. We obtain somewhat high values of $b_3$ at low $b_1$ (see \reffig{b3}) which could explain why we find high values for $b_2$ as well.

\begin{figure}
\centering
\includegraphics[scale=0.28]{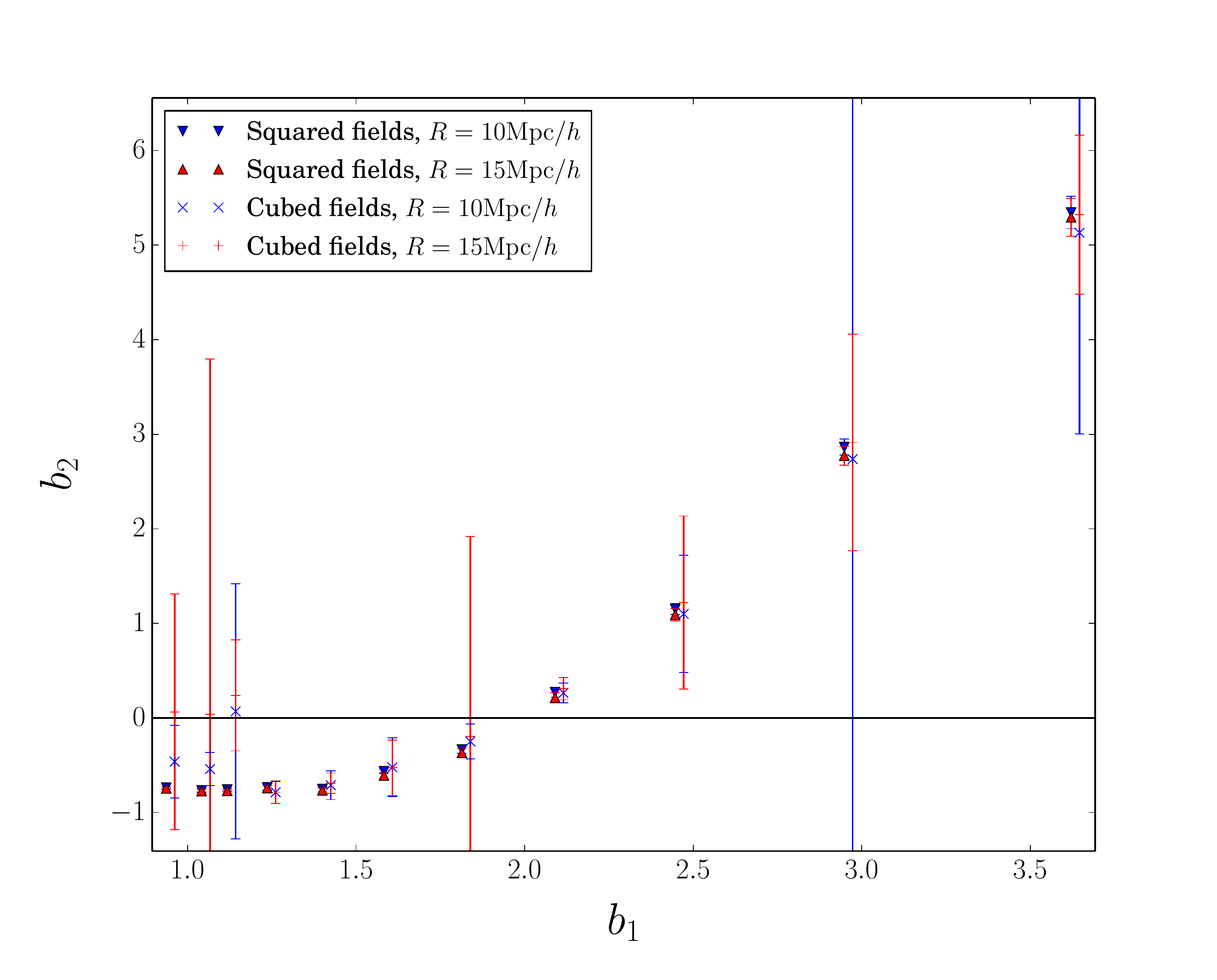}
\includegraphics[scale=0.28]{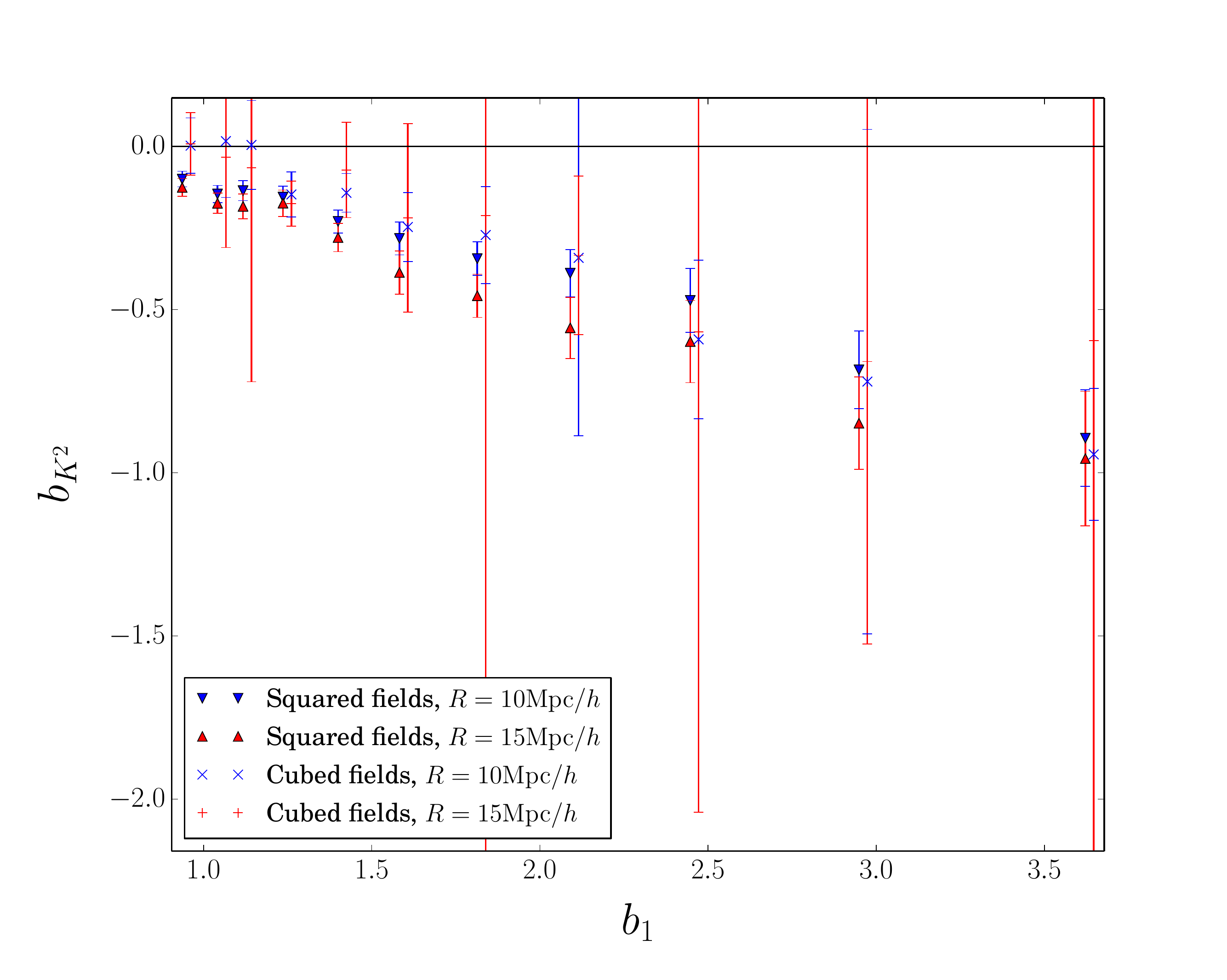}
\caption{$b_2$ (left) and $b_{K^2}$ (right) as a function of $b_1$ for two different smoothing scales $R$. The color coding is the same as in \reffig{kdep} as well as the horizontal shift for the cubed-fields results. Again the results within each method are self-consistent, proving the robustness of the results under a change in $R$. }
\label{fig:Rdep}
\end{figure}

\refFig{Rdep} presents a similar comparison, but now for two different smoothing scales $R$, namely $R=10$ and $15 \Mpch$. 
The conclusions are the same as for \reffig{kdep}, i.e. good agreement between results within each method, but we observe the same lack of signal at both smoothing scales for the cubed-field results from the L500 set. 
As we expected, the results do not depend strongly on the choice of $R$, but the constraining power increases for smaller $R$. We chose to use $R=15\Mpch$ for the final results, since we expect nonlinearities to begin to have a significant impact for $R=10\Mpch$, while the constraining power dramatically weakens for larger values of $R$. 

\begin{figure}
\centering
\includegraphics[scale=0.28]{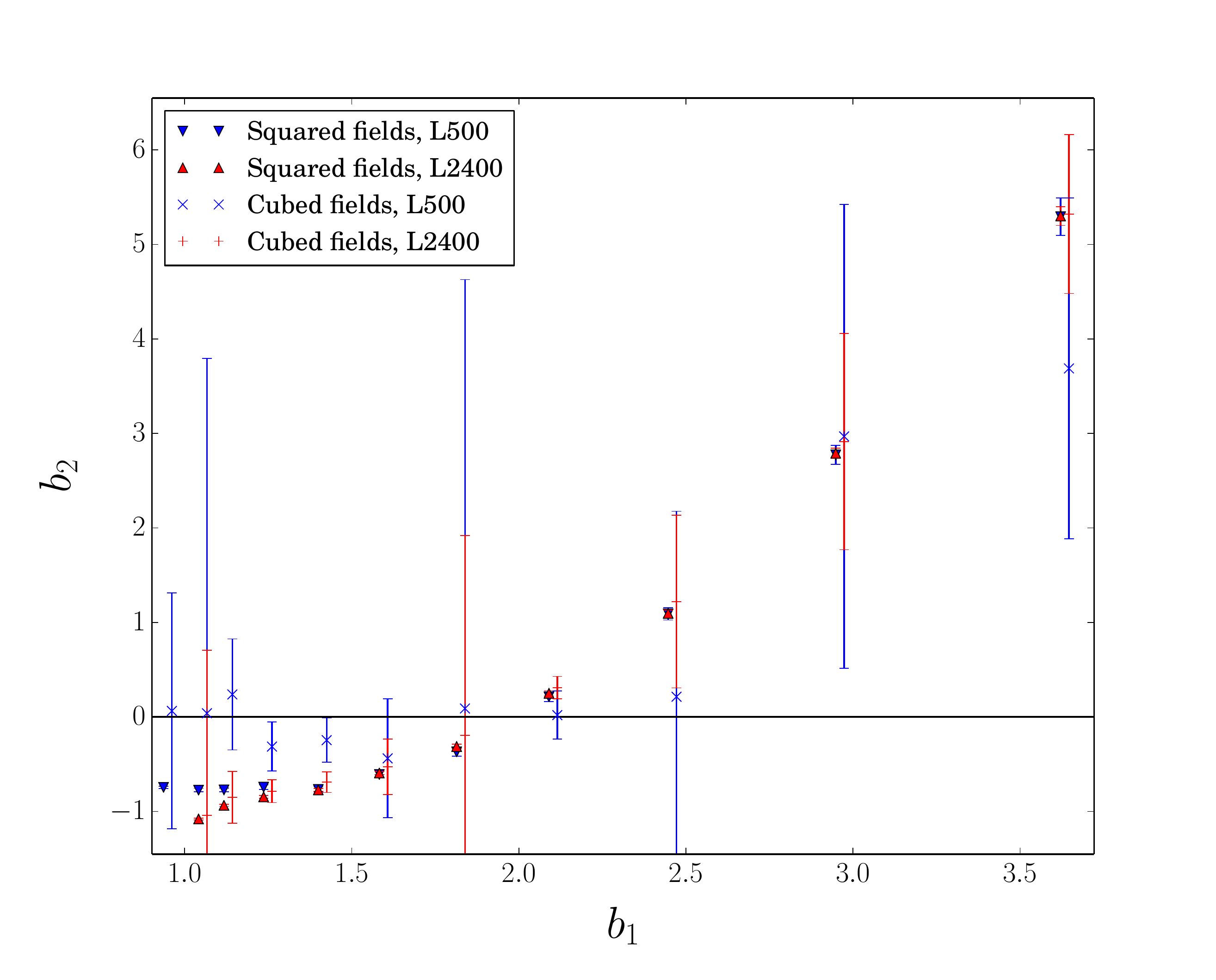}
\includegraphics[scale=0.28]{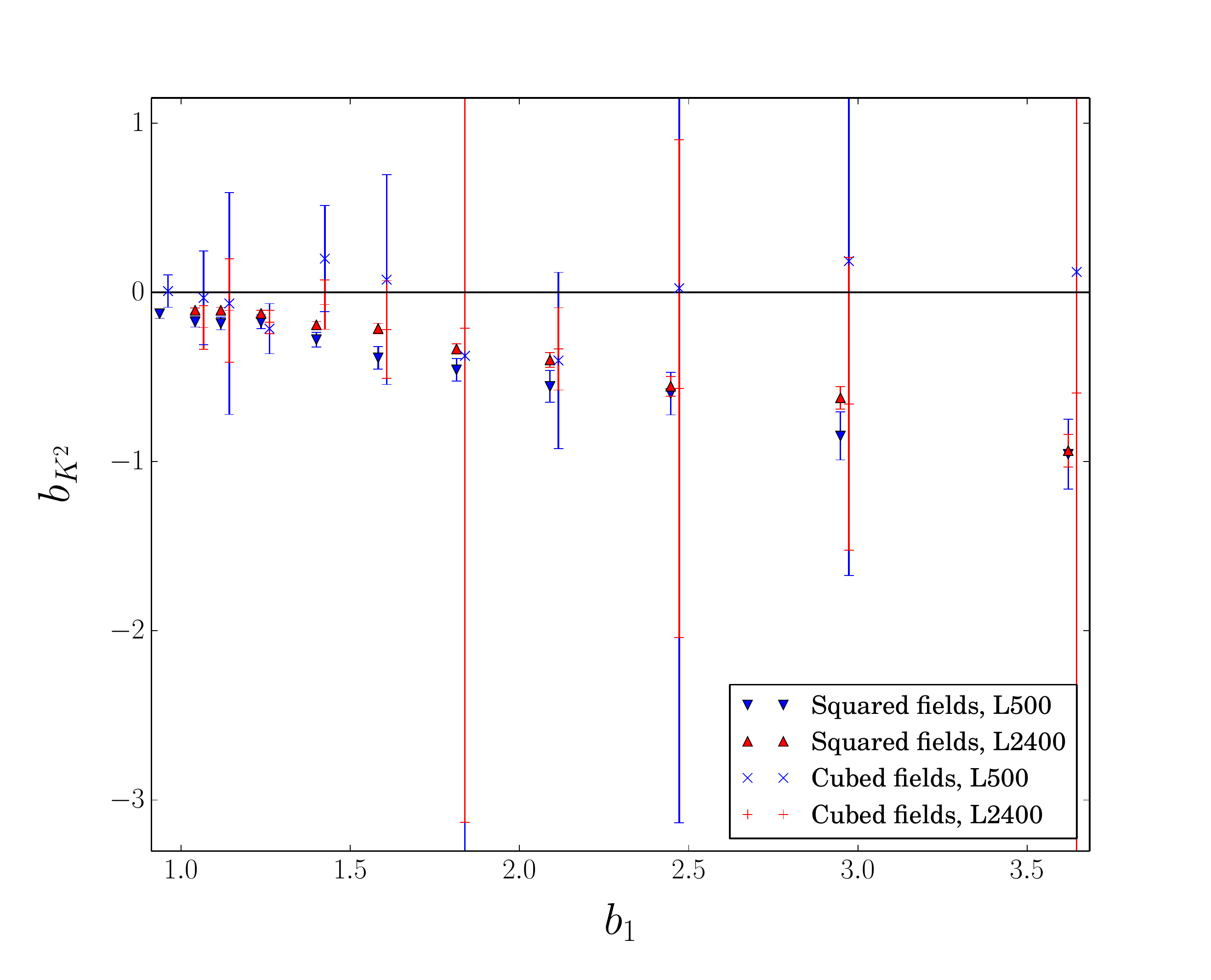}
\caption{$b_2$ (left) and $b_{K^2}$ (right) as a function of $b_1$ for the two different simulations sets, with the color coding again following that of \reffig{kdep} and the cubed-fields points being slightly displaced horizontally for clarity. The findings are the same as for the previous checks: the results do not depend significantly on the simulation set used for each method, but a lack of signal is observed for results from cubed-fields using the L500 set}.
\label{fig:Ldep}
\end{figure}

Finally, \reffig{Ldep} presents the comparison of the results obtained with each simulation set for $R=15 \Mpch$ and $k=0.18$ Mpc$/h$. We again see a very good agreement between the two sets of simulations for each method, and a low signal-to-noise ratio for the cubed-fields results using the L500 set. We insist that this lack of signal is only observed in the results for the quadratic parameters. This justifies our use of the L2400 set for bins at sufficiently high mass (in order to maximize the constraining power) and the L500 set to push down to lower mass.

Finally, we briefly address the large fluctuations of the errorbars across $b_1$ that can be seen in our results. We have derived the covariance of our results between mass bins and found it to be rather small overall, which explains why these fluctuations are possible. One possible reason for their origin is the fact that we invert the matrix $M_{OO'}$ to obtain results for $\v{c}$ which leads to a nontrivial propagation of errors. However, these fluctuations do not affect our overall results and conclusions.
 
The results of this appendix show the robustness of our results under the change of various parameters and motivate our choices for the final measurements. 

\section{Covariance matrix}
\label{app:covariance}

This section presents the covariance matrix of the parameter combinations entering \refeq{C}. We present here the correlation coefficient corresponding to the inverse of the sample covariance obtained from the 48 realisations of the L500 set. The correlation coefficient is defined as
\be
\rho_{OO'}=\frac{C_{OO'}}{\sqrt{C_{OO}C_{O'O'}}} \, ,
\label{eq:corrcoeff}
\ee
where
  \be
  C_{OO'} = \< c_O c_{O'} \> - \< c_O \> \< c_{O'} \>
  \ee
  is the covariance matrix of the parameters $\{c_O\}$, and the expectation value is over simulation realizations. Note that both $C_{OO'}$ and $\rho_{OO'}$ refer to the parameters $c_O$, rather than operators $O$. Notice also that our bootstrap technique yields the error bars on the $c_{O}$ after marginalization over all other $c_{O'}$, i.e.
\be
\sigma_\text{marg}(c_O) = \left(C_{OO}\right)^{1/2} \, .
\ee

\begin{figure}
\centering
\includegraphics[scale=0.37]{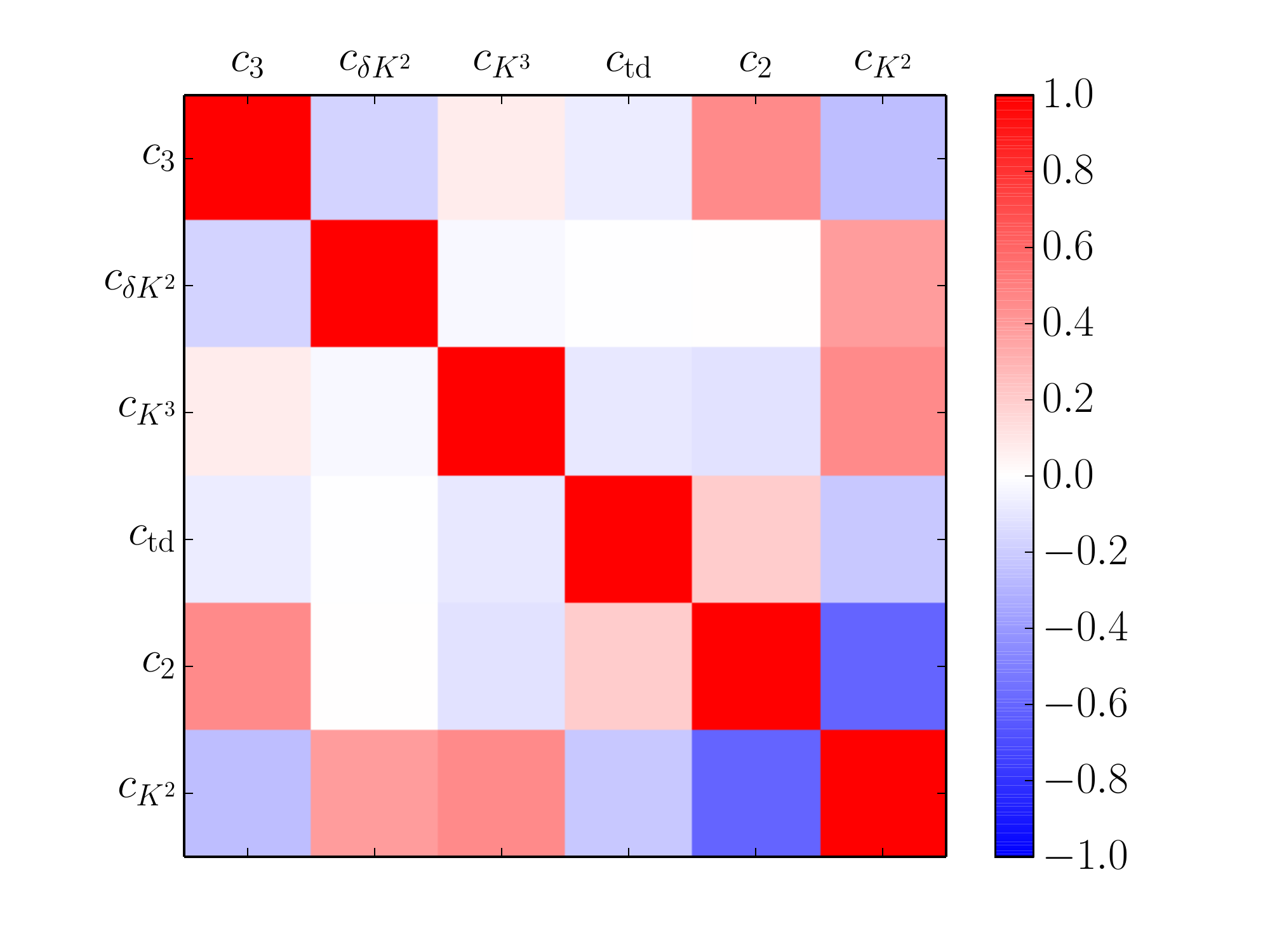}
\includegraphics[scale=0.37]{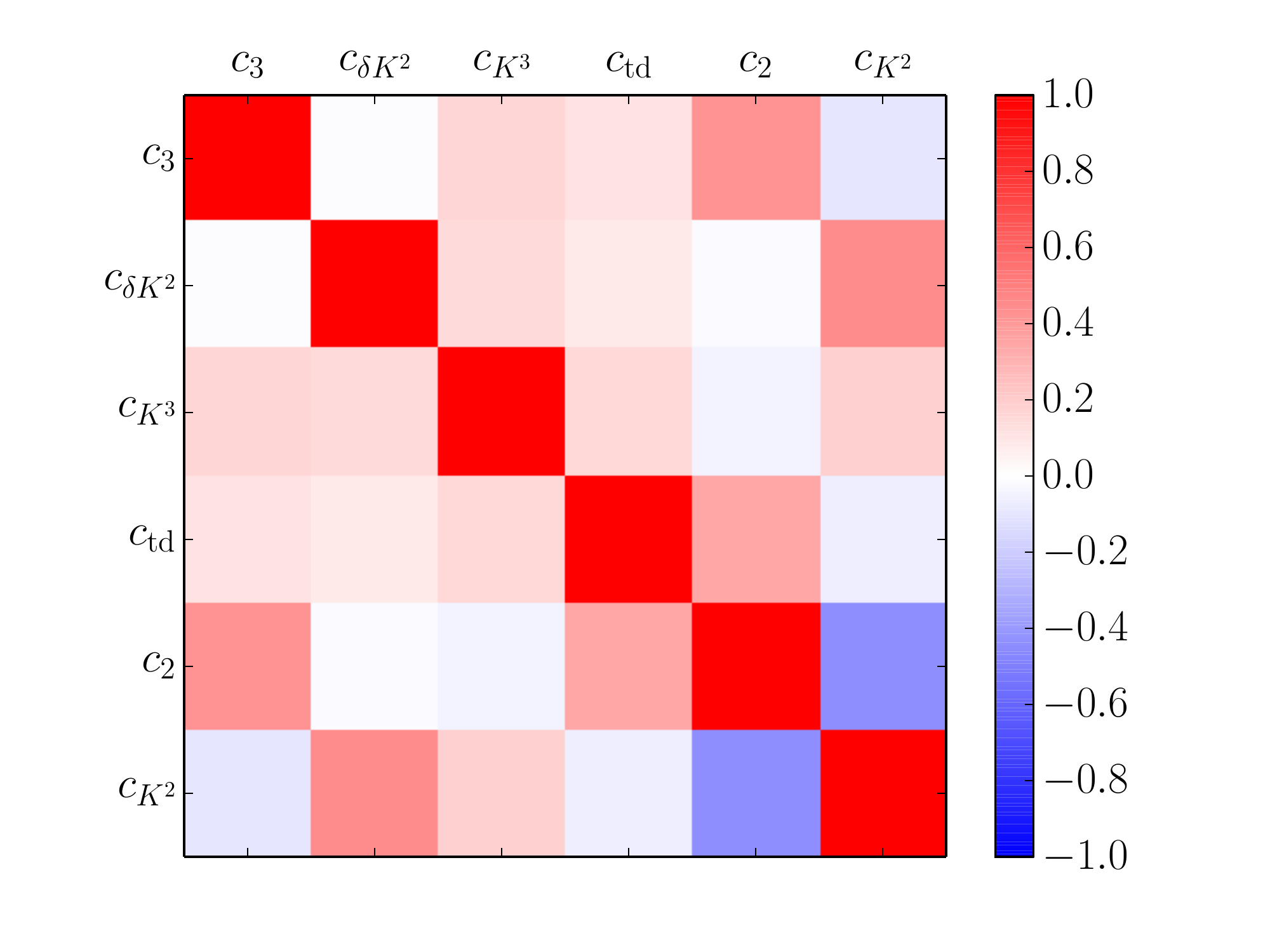}
\caption{The correlation coefficient corresponding to \refeq{corrcoeff} of the parameters combinations entering \refeq{C} for the two mass bins $\lg M=12.55 \mbox{ (left) and } 13.95 \, {\rm M}_\odot/h$ (right). We use the notation $c_O$ for the elements of $\v{c}$ where $O$ is the subscript of the first parameter entering each combination in \refeq{C}. A white cell indicate zero correlation, whilst blue and red cells indicate negative and positive correlations respectively. We discuss these correlations in more details in the text.}
\label{fig:cova}
\end{figure}

Since the covariance is stable through all mass bins, we only show results for a couple of representative mass bins $\lg M=12.55 \mbox{ and } 13.95 \, {\rm M}_\odot/h$ in \reffig{cova}. We use the notation $c_O$ for the elements of $\v{c}$ where $O$ is the subscript of the first parameter entering each combination in \refeq{C}. The correlation coefficient is under 0.2 between most combinations, indicating only low covariance. It can however be important (of the order of 0.5) between $c_2$ or $c_{K^2}$ and other $c_O$. It is maximal between $c_2$ and $c_3$, and between $c_{K^2}$ and $c_{\d K^2}$, and can also be important between $c_2$ and $c_{K^2}$ as well as between $c_{K^2}$ and $c_{K^3}$.

\section{Consistency checks: $b_1$, $b_2$ and $b_3$}
\label{app:check}

In this appendix we present our results for the LIMD bias parameters up to cubic order at redshift 0. These parameters have already been studied quite extensively in the literature (especially $b_1$ and $b_2$). Here, we compare our results to those of L15 as a check of our method. We use the separate-universe measurements of L15, as they have been obtained for the exact same cosmology and for a similar set of simulations. We refer the reader to L15 for a detailed comparison of these parameters with previous results and various analytical predictions.

\begin{figure}
\centering
\includegraphics[scale=0.42]{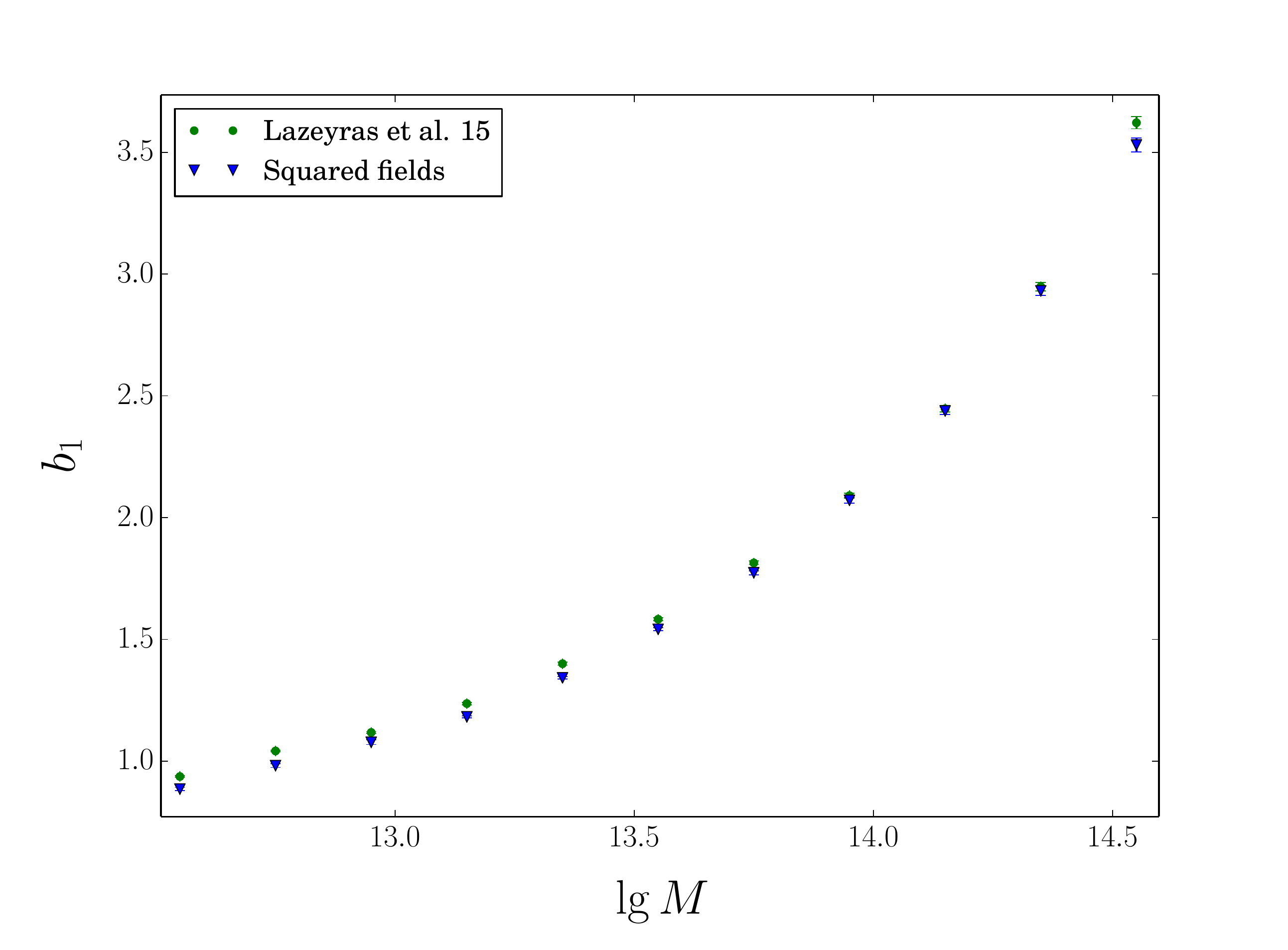}
\caption{$b_1$ as a function of $\lg M$. The green dots are the results of L15 while the blue triangles are the results obtained from squared-field correlators. Although there seems to be a small systematic shift for lower masses the overall agreement is very satisfying.}
\label{fig:b1}
\end{figure}

We start by comparing the results for $b_1$. As can be seen in \reffig{b1}, although there seems to be a small systematic shift between our results from squared-field correlators and those of L15 for lower masses, the overall agreement is very satisfying and provides a good first validation of our method. It is also worth noticing that the error bars are of roughly the same size for both measurement sets.

\begin{figure}
\centering
\includegraphics[scale=0.4]{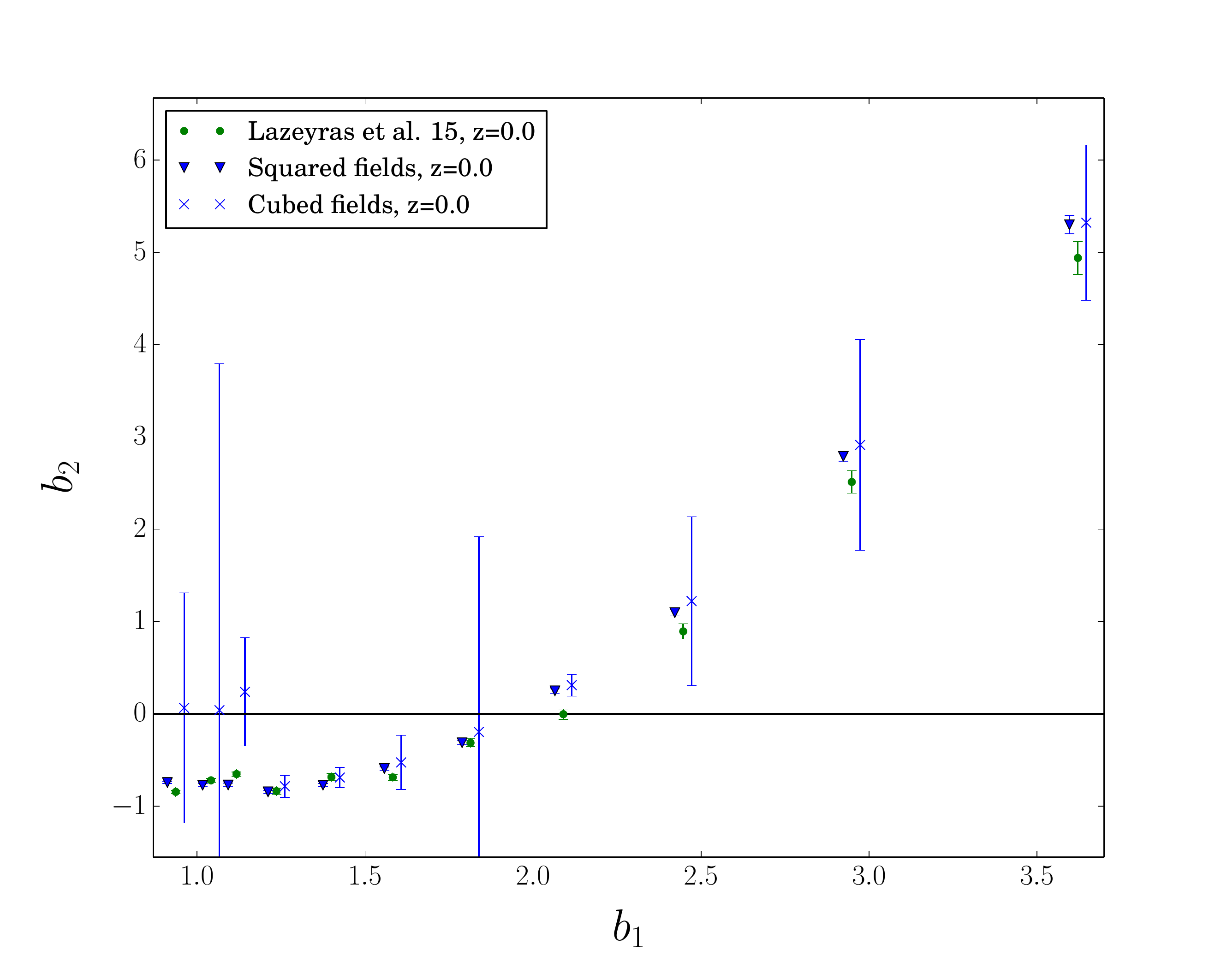}
\caption{$b_2$ as a function of $b_1$. The color coding is the same as in \reffig{b1}, but we now also show measurements from correlators of cubed fields, denoted by blue crosses. The overall agreement of both sets with the L15 results is good, although the low signal-to-noise ratio for cubed-fields results discussed in \refapp{convergence} is evident.}
\label{fig:b2}
\end{figure}

\refFig{b2} presents the results for $b_2$, for which we have measurements both from squared and cubed-field correlators. The overall agreement of both sets with the L15 results is good, although the cubed-field result shows a low signal-to-noise ratio for the three lowest mass bins as discussed in \refapp{convergence}. Nevertheless, for $b_1 \gtrsim 1.25$, the cubed-field result for $b_2$ is consistent within errors with the separate-universe measurements from L15.

\begin{figure}
\centering
\includegraphics[scale=0.4]{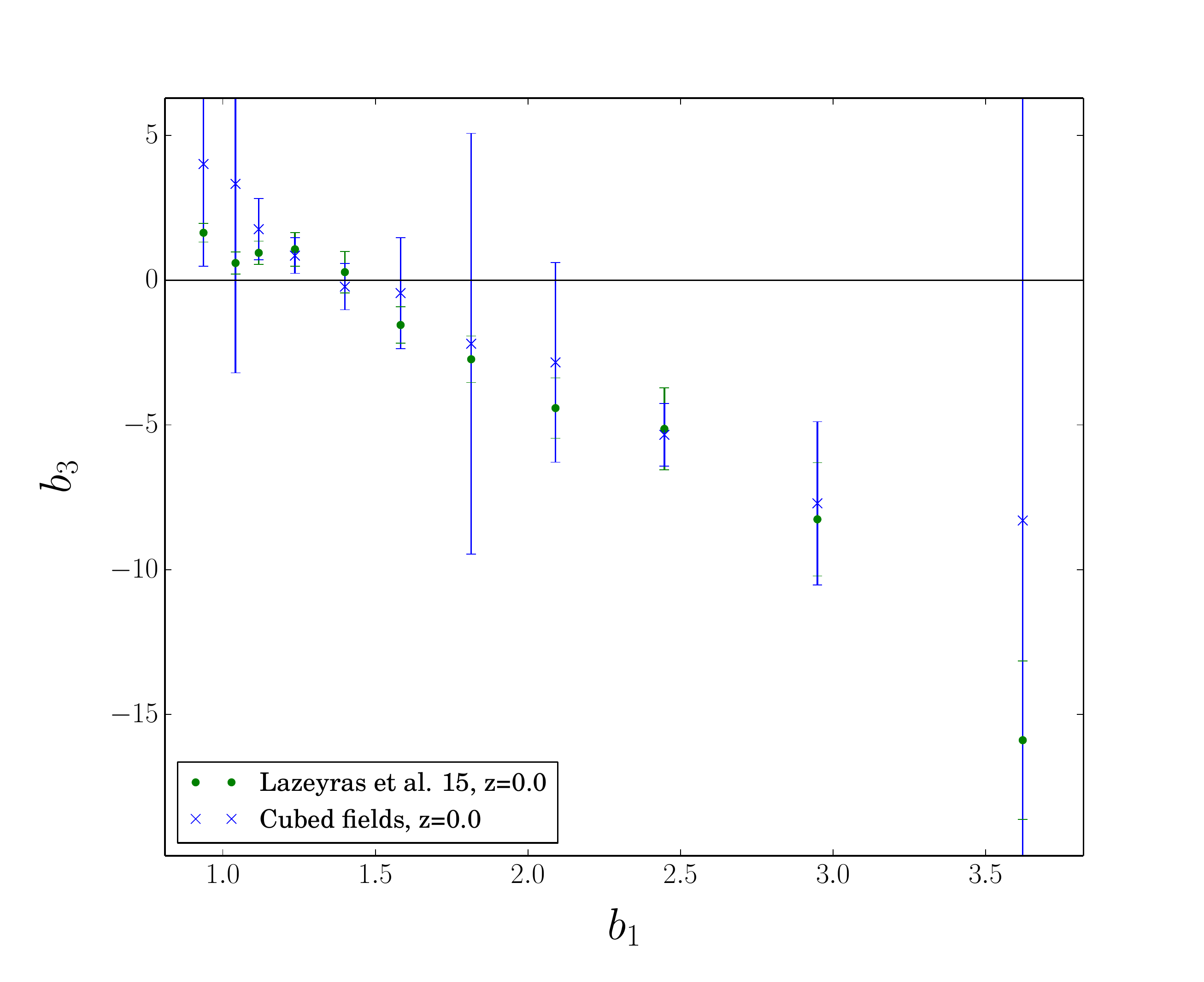}
\caption{$b_3$ as a function of $b_1$. The color coding is the same as in \reffig{b2}. The agreement between the two measurement methods for this bias parameter is excellent.}
\label{fig:b3}
\end{figure}

Finally, \reffig{b3} presents the comparison between our measurements from cubed-field correlators and the L15 results. The overall agreement between the two sets is excellent, with only a few mass bins showing some mildly discrepant values. 


\FloatBarrier

\bibliography{references1}

\providecommand{\href}[2]{#2}\begingroup\raggedright\begin{thebibliography}{10}

\bibitem{Abidi:2018eyd}
M.~M. Abidi and T.~Baldauf, {\it {Cubic Halo Bias in Eulerian and Lagrangian
  Space}},  \href{http://arxiv.org/abs/1802.07622}{{\tt arXiv:1802.07622}}.

\bibitem{Brainerd:1995da}
T.~G. Brainerd, R.~D. Blandford, and I.~Smail, {\it {Measuring galaxy masses
  using galaxy - galaxy gravitational lensing}},  {\em Astrophys. J.} {\bf 466}
  (1996) 623, [\href{http://arxiv.org/abs/astro-ph/9503073}{{\tt
  astro-ph/9503073}}].

\bibitem{biasreview}
V.~Desjacques, D.~Jeong, and F.~Schmidt, {\it {Large-Scale Galaxy Bias}},
  \href{http://arxiv.org/abs/1611.09787}{{\tt arXiv:1611.09787}}.

\bibitem{senatore:2015}
L.~{Senatore}, {\it {Bias in the effective field theory of large scale
  structures}},  {\em \jcap} {\bf 11} (Nov., 2015) 007,
  [\href{http://arxiv.org/abs/1406.7843}{{\tt arXiv:1406.7843}}].

\bibitem{MSZ}
M.~{Mirbabayi}, F.~{Schmidt}, and M.~{Zaldarriaga}, {\it {Biased tracers and
  time evolution}},  {\em \jcap} {\bf 7} (July, 2015) 30,
  [\href{http://arxiv.org/abs/1412.5169}{{\tt arXiv:1412.5169}}].

\bibitem{baldauf/etal:2011}
T.~{Baldauf}, U.~{Seljak}, L.~{Senatore}, and M.~{Zaldarriaga}, {\it {Galaxy
  bias and non-linear structure formation in general relativity}},  {\em \jcap}
  {\bf 10} (Oct., 2011) 031, [\href{http://arxiv.org/abs/1106.5507}{{\tt
  arXiv:1106.5507}}].

\bibitem{CFCpaper1}
L.~{Dai}, E.~{Pajer}, and F.~{Schmidt}, {\it {Conformal Fermi Coordinates}},
  {\em \jcap} {\bf 11} (Nov., 2015) 043,
  [\href{http://arxiv.org/abs/1502.02011}{{\tt arXiv:1502.02011}}].

\bibitem{mcdonald/roy}
P.~{McDonald} and A.~{Roy}, {\it {Clustering of dark matter tracers:
  generalizing bias for the coming era of precision LSS}},  {\em \jcap} {\bf 8}
  (Aug., 2009) 20, [\href{http://arxiv.org/abs/0902.0991}{{\tt
  arXiv:0902.0991}}].

\bibitem{assassi/etal}
V.~{Assassi}, D.~{Baumann}, D.~{Green}, and M.~{Zaldarriaga}, {\it
  {Renormalized halo bias}},  {\em \jcap} {\bf 8} (Aug., 2014) 56,
  [\href{http://arxiv.org/abs/1402.5916}{{\tt arXiv:1402.5916}}].

\bibitem{angulo/etal}
R.~{Angulo}, M.~{Fasiello}, L.~{Senatore}, and Z.~{Vlah}, {\it {On the
  statistics of biased tracers in the Effective Field Theory of Large Scale
  Structures}},  {\em \jcap} {\bf 9} (Sept., 2015) 029,
  [\href{http://arxiv.org/abs/1503.08826}{{\tt arXiv:1503.08826}}].

\bibitem{Matarrese:1997sk}
S.~Matarrese, L.~Verde, and A.~F. Heavens, {\it {Large scale bias in the
  universe: Bispectrum method}},  {\em Mon. Not. Roy. Astron. Soc.} {\bf 290}
  (1997) 651--662, [\href{http://arxiv.org/abs/astro-ph/9706059}{{\tt
  astro-ph/9706059}}].

\bibitem{Angulo:2007}
R.~E. Angulo, C.~M. Baugh, and C.~G. Lacey, {\it {The assembly bias of dark
  matter haloes to higher orders}},  {\em Mon. Not. Roy. Astron. Soc.} {\bf
  387} (2008) 921, [\href{http://arxiv.org/abs/0712.2280}{{\tt
  arXiv:0712.2280}}].

\bibitem{guo/jing:2009}
H.~{Guo} and Y.~P. {Jing}, {\it {Determine the Galaxy Bias Factors on Large
  Scales Using the Bispectrum Method}},  {\em \apj} {\bf 702} (Sept., 2009)
  425--432, [\href{http://arxiv.org/abs/0907.0282}{{\tt arXiv:0907.0282}}].

\bibitem{tinker/etal:2010}
J.~L. {Tinker}, B.~E. {Robertson}, A.~V. {Kravtsov}, A.~{Klypin}, M.~S.
  {Warren}, G.~{Yepes}, and S.~{Gottl{\"o}ber}, {\it {The Large-scale Bias of
  Dark Matter Halos: Numerical Calibration and Model Tests}},  {\em \apj} {\bf
  724} (Dec., 2010) 878--886, [\href{http://arxiv.org/abs/1001.3162}{{\tt
  arXiv:1001.3162}}].

\bibitem{pollack/smith/porciani:2012}
J.~E. {Pollack}, R.~E. {Smith}, and C.~{Porciani}, {\it {Modelling large-scale
  halo bias using the bispectrum}},  {\em \mnras} {\bf 420} (Mar., 2012)
  3469--3489, [\href{http://arxiv.org/abs/1109.3458}{{\tt arXiv:1109.3458}}].

\bibitem{Saito:2014qha}
S.~Saito, T.~Baldauf, Z.~Vlah, U.~Seljak, T.~Okumura, and P.~McDonald, {\it
  {Understanding higher-order nonlocal halo bias at large scales by combining
  the power spectrum with the bispectrum}},  {\em Phys. Rev.} {\bf D90} (2014),
  no.~12 123522, [\href{http://arxiv.org/abs/1405.1447}{{\tt
  arXiv:1405.1447}}].

\bibitem{Lazeyras:2015}
T.~Lazeyras, C.~Wagner, T.~Baldauf, and F.~Schmidt, {\it {Precision measurement
  of the local bias of dark matter halos}},  {\em JCAP} {\bf 1602} (2016),
  no.~02 018, [\href{http://arxiv.org/abs/1511.01096}{{\tt arXiv:1511.01096}}].

\bibitem{Li:2015}
Y.~{Li}, W.~{Hu}, and M.~{Takada}, {\it {Separate Universe Consistency Relation
  and Calibration of Halo Bias}},  {\em ArXiv e-prints} (Nov., 2015)
  [\href{http://arxiv.org/abs/1511.01454}{{\tt arXiv:1511.01454}}].

\bibitem{Baldauf:2015}
T.~{Baldauf}, U.~{Seljak}, L.~{Senatore}, and M.~{Zaldarriaga}, {\it {Linear
  response to long wavelength fluctuations using curvature simulations}},  {\em
  ArXiv e-prints} (Nov., 2015) [\href{http://arxiv.org/abs/1511.01465}{{\tt
  arXiv:1511.01465}}].

\bibitem{hoffmann/etal:2015}
K.~{Hoffmann}, J.~{Bel}, E.~{Gazta{\~n}aga}, M.~{Crocce}, P.~{Fosalba}, and
  F.~J. {Castander}, {\it {Measuring the growth of matter fluctuations with
  third-order galaxy correlations}},  {\em \mnras} {\bf 447} (Feb., 2015)
  1724--1745, [\href{http://arxiv.org/abs/1403.1259}{{\tt arXiv:1403.1259}}].

\bibitem{Hoffmann:2016omy}
K.~Hoffmann, J.~Bel, and E.~Gaztanaga, {\it {Linear and non-linear bias:
  predictions versus measurements}},  {\em Mon. Not. Roy. Astron. Soc.} {\bf
  465} (2017), no.~2 2225--2235, [\href{http://arxiv.org/abs/1607.01024}{{\tt
  arXiv:1607.01024}}].

\bibitem{manera/gaztanaga:2011}
M.~{Manera} and E.~{Gazta{\~n}aga}, {\it {The local bias model in the
  large-scale halo distribution}},  {\em \mnras} {\bf 415} (July, 2011)
  383--398, [\href{http://arxiv.org/abs/0912.0446}{{\tt arXiv:0912.0446}}].

\bibitem{castorina/paranjape/etal:2016}
E.~{Castorina}, A.~{Paranjape}, O.~{Hahn}, and R.~K. {Sheth}, {\it {Excursion
  set peaks: the role of shear}},  {\em ArXiv e-prints} (Nov., 2016)
  [\href{http://arxiv.org/abs/1611.03619}{{\tt arXiv:1611.03619}}].

\bibitem{McDonald:2001fe}
P.~McDonald, {\it {Toward a measurement of the cosmological geometry at Z 2:
  predicting lyman-alpha forest correlation in three dimensions, and the
  potential of future data sets}},  {\em Astrophys. J.} {\bf 585} (2003)
  34--51, [\href{http://arxiv.org/abs/astro-ph/0108064}{{\tt
  astro-ph/0108064}}].

\bibitem{Wagner:2014}
C.~Wagner, F.~Schmidt, C.-T. Chiang, and E.~Komatsu, {\it {Separate Universe
  Simulations}},  {\em Mon.Not.Roy.Astron.Soc.} {\bf 448} (2015) 11,
  [\href{http://arxiv.org/abs/1409.6294}{{\tt arXiv:1409.6294}}].

\bibitem{Schmittfull:2014tca}
M.~Schmittfull, T.~Baldauf, and U.~Seljak, {\it {Near optimal bispectrum
  estimators for large-scale structure}},  {\em Phys. Rev.} {\bf D91} (2015),
  no.~4 043530, [\href{http://arxiv.org/abs/1411.6595}{{\tt arXiv:1411.6595}}].

\bibitem{chan/scoccimarro/sheth:2012}
K.~C. {Chan}, R.~{Scoccimarro}, and R.~K. {Sheth}, {\it {Gravity and
  large-scale nonlocal bias}},  {\em \prd} {\bf 85} (Apr., 2012) 083509,
  [\href{http://arxiv.org/abs/1201.3614}{{\tt arXiv:1201.3614}}].

\bibitem{baldauf/etal:2012}
T.~{Baldauf}, U.~{Seljak}, V.~{Desjacques}, and P.~{McDonald}, {\it {Evidence
  for quadratic tidal tensor bias from the halo bispectrum}},  {\em \prd} {\bf
  86} (Oct., 2012) 083540, [\href{http://arxiv.org/abs/1201.4827}{{\tt
  arXiv:1201.4827}}].

\bibitem{sheth/chan/scoccimarro:2012}
R.~K. {Sheth}, K.~C. {Chan}, and R.~{Scoccimarro}, {\it {Nonlocal Lagrangian
  bias}},  {\em \prd} {\bf 87} (Apr., 2013) 083002,
  [\href{http://arxiv.org/abs/1207.7117}{{\tt arXiv:1207.7117}}].

\bibitem{bel/hoffmann/gaztanaga:15}
J.~{Bel}, K.~{Hoffmann}, and E.~{Gazta{\~n}aga}, {\it {Non-local bias
  contribution to third-order galaxy correlations}},  {\em \mnras} {\bf 453}
  (Oct., 2015) 259--276, [\href{http://arxiv.org/abs/1504.02074}{{\tt
  arXiv:1504.02074}}].

\bibitem{Modi:2016dah}
C.~Modi, E.~Castorina, and U.~Seljak, {\it {Halo bias in Lagrangian Space:
  Estimators and theoretical predictions}},  {\em Mon. Not. Roy. Astron. Soc.}
  {\bf 472} (2017) 3959, [\href{http://arxiv.org/abs/1612.01621}{{\tt
  arXiv:1612.01621}}].

\bibitem{Springel:2005}
V.~Springel, {\it {The Cosmological simulation code GADGET-2}},  {\em
  Mon.Not.Roy.Astron.Soc.} {\bf 364} (2005) 1105--1134,
  [\href{http://arxiv.org/abs/astro-ph/0505010}{{\tt astro-ph/0505010}}].

\bibitem{Gill:2004}
S.~P. Gill, A.~Knebe, and B.~K. Gibson, {\it {The Evolution substructure 1: A
  New identification method}},  {\em Mon.Not.Roy.Astron.Soc.} {\bf 351} (2004)
  399, [\href{http://arxiv.org/abs/astro-ph/0404258}{{\tt astro-ph/0404258}}].

\bibitem{Knollmann:2009}
S.~R. Knollmann and A.~Knebe, {\it {Ahf: Amiga's Halo Finder}},  {\em
  Astrophys.J.Suppl.} {\bf 182} (2009) 608--624,
  [\href{http://arxiv.org/abs/0904.3662}{{\tt arXiv:0904.3662}}].

\bibitem{catelan/etal:1998}
P.~{Catelan}, F.~{Lucchin}, S.~{Matarrese}, and C.~{Porciani}, {\it {The bias
  field of dark matter haloes}},  {\em \mnras} {\bf 297} (July, 1998) 692--712,
  [\href{http://arxiv.org/abs/astro-ph/9}{{\tt astro-ph/9}}].

\bibitem{catelan/porciani/kamionkowski:2000}
P.~{Catelan}, C.~{Porciani}, and M.~{Kamionkowski}, {\it {Two ways of biasing
  galaxy formation}},  {\em \mnras} {\bf 318} (Nov., 2000) L39--L44,
  [\href{http://arxiv.org/abs/astro-ph/0}{{\tt astro-ph/0}}].

\bibitem{Fujita:2016}
T.~Fujita, V.~Mauerhofer, L.~Senatore, Z.~Vlah, and R.~Angulo, {\it {Very
  Massive Tracers and Higher Derivative Biases}},
  \href{http://arxiv.org/abs/1609.00717}{{\tt arXiv:1609.00717}}.

\bibitem{mcdonald}
P.~{McDonald}, {\it {Clustering of dark matter tracers: Renormalizing the bias
  parameters}},  {\em \prd} {\bf 74} (Nov., 2006) 103512,
  [\href{http://arxiv.org/abs/astro-ph/0}{{\tt astro-ph/0}}].

\bibitem{PBSpaper}
F.~{Schmidt}, D.~{Jeong}, and V.~{Desjacques}, {\it {Peak-background split,
  renormalization, and galaxy clustering}},  {\em \prd} {\bf 88} (July, 2013)
  023515, [\href{http://arxiv.org/abs/1212.0868}{{\tt arXiv:1212.0868}}].

\end{thebibliography}\endgroup
\end{document}